\documentclass[aps,prb,reprint,twocolumn,superscriptaddress,floatfix,longbibliography]{revtex4-1}
\PassOptionsToPackage{usenames,dvipsnames}{xcolor}

\usepackage{graphicx}
\usepackage{amsmath,amsfonts,amssymb,amsthm}
\usepackage{mathtools}
\usepackage{mathrsfs}
\usepackage{bbold}
\usepackage{dsfont}
\usepackage{physics}

\usepackage{tikz}

\usepackage[english]{babel}
\makeatletter
\renewcommand{\selectlanguage}[1]{}
\makeatother

\usepackage{xcolor}
\usepackage[bookmarks=true,colorlinks,linkcolor=OrangeRed,urlcolor=NavyBlue,citecolor=RoyalBlue]{hyperref}

\usepackage{booktabs}
\usepackage{siunitx}
\usepackage{microtype}
\usepackage{orcidlink}

\definecolor{mygold}{rgb}{0.93,0.69,0.13}
\definecolor{mypurple}{rgb}{0.49,0.18,0.56}
\definecolor{mygreen}{rgb}{0,0.5,0}
\definecolor{myred}{rgb}{0.7,0,0}
\definecolor{myblue}{rgb}{0,0,1}

\newcommand{\pihat}[2]{\mathop{\hat{\Pi}}\limits_{\scriptstyle #1}^{\scriptstyle #2}}
\begin{document}
\title{Matrix Product Operator Constructions for Gauge Theories in the Thermodynamic Limit}

\author{Nicholas Godfrey}
\affiliation{Queensland University of Technology (QUT)}
\affiliation{School of Mathematics and Physics, University of Queensland, St Lucia, Queensland 4072, Australia}

\author{Ian P.~McCulloch${}^{\orcidlink{0000-0002-8983-6327}}$}
\affiliation{Frontier Center for Theory and Computation, National Tsing Hua University, Hsinchu 30013, Taiwan}
\affiliation{Department of Physics, National Tsing Hua University, Hsinchu 30013, Taiwan}
\affiliation{School of Maths and Physics, University of Queensland, St Lucia, Queensland 4072, Australia}

\begin{abstract}
  We present a general method for simulating lattice gauge theories in low dimensions using infinite matrix product states (iMPS). A central challenge in Hamiltonian formulations of gauge theories is the unbounded local Hilbert space associated with gauge degrees of freedom. In one spatial dimension, Gauss’s law permits these gauge fields to be integrated out, yielding an effective Hamiltonian with long-range interactions among matter fields. We construct efficient matrix product operator (MPO) representations of these Hamiltonians directly in the thermodynamic limit. Our formulation naturally includes background fields and $\theta$-terms, requiring no modifications to the standard iDMRG algorithm. This provides a broadly applicable framework for 1+1D gauge theories and can be extended to quasi-two-dimensional geometries such as infinite cylinders, where tensor-network methods remain tractable. As a benchmark, we apply our construction to the Schwinger model, reproducing expected features including confinement, string breaking, and the critical behavior at finite mass. Because the method alters only the MPO structure, it can be incorporated with little effort into a wide range of iMPS and infinite-boundary-condition algorithms, opening the way to efficient studies of both equilibrium and non-equilibrium gauge dynamics.
\end{abstract}

\maketitle

\section{Introduction}
Gauge theories play a central role in both high-energy and condensed matter physics, providing the framework for understanding phenomena such as confinement and fractionalization. A paradigmatic example is the Schwinger model, which describes quantum electrodynamics in one spatial dimension $(1+1D)$ with charged fermions and a $U(1)$ gauge field. Despite its relative simplicity, the Schwinger model captures essential non-perturbative physics: for instance, the linear confinement of electric flux tubes and their decay via string breaking, where a strong electric field between two charges can produce particle–antiparticle pairs. Such processes are not only fundamental in particle physics (e.g. modeling aspects of quark confinement in QCD) but also have analogues in strongly correlated condensed matter systems\cite{LiuStringBreaking}. This has spurred interest in alternative computational and experimental methods to tackle non-perturbative gauge physics. The Schwinger model serves as a valuable testbed for this, as it is relatively straightforward to solve numerically. In the typical lattice discretization it is defined on a one-dimensional lattice with staggered fermions and gauge links\cite{kogut_hamiltonian_1975}, and can include a background electric field or topological $\theta$-term.

Conventional Monte Carlo approaches have achieved spectacular success for static observables of lattice gauge theories, but they face fundamental obstacles in addressing real-time dynamics, finite density, and topological terms. These limitations have motivated the exploration of alternative frameworks, ranging from quantum simulation experiments to tensor-network methods. In particular, tensor networks offer a controlled, variational approach to gauge theories in low dimensions that naturally avoids the sign problem and allows access to both ground-state and dynamical properties.
Building on early density matrix renormalization group (DMRG) treatments of the massive Schwinger model \cite{byrnes_density_2002,byrnes_density_2003}, matrix product state (MPS) methods have been developed into a versatile tool for gauge theories. Applications include the determination of the mass spectrum and excitations \cite{banuls_mass_2013}, studies of density-induced phase transitions \cite{banuls_density_2017}, and real-time dynamics such as Schwinger pair production \cite{buyens_real-time_2017}. Extensions have also addressed finite temperature, finite density, and continuum extrapolations \cite{buyens_finite-representation_2017,papaefstathiou_density_2021}. Recent reviews summarize these advances and their prospects for higher-dimensional gauge theories\cite{banuls_review_2020,Magnifico2025TensorNetworks,KelmanGaussian2024,Montangero2021LoopFreeTensor,Zohar2022QuantumSimulationLGTs,Aidelsburger2022ColdAtomsLGT}. These works firmly establish tensor networks as a benchmark numerical approach for non-perturbative gauge dynamics.

A complication with lattice gauge theories is the gauge degrees of freedom have an unbounded Hilbert space, and hence in practice one often resorts to approximations. One approach is to modify the algebra of the gauge particles to be finite (resulting in a so-called quantum link model). These models are interesting in their own right, but they are only approximations to the actual gauge theory. For simulating the gauge theory itself in numerical approaches, the maximum boson number can be truncated\cite{buyens_finite-representation_2017}; if this is done at a sufficiently large number then it should have negligible effect on the results. However this can  significantly reduce the efficiency. With some effort it is possible to construct the DMRG algorithm so that it is linear in the local Hilbert space dimension\cite{PhysRevLett.130.246402,CBEComment}, but conventional algorithms have  quadratic or worse scaling. As a special case, in one dimension it is possible to integrate out the gauge degrees of freedom. Since the flux lines connecting particle-antiparticle pairs are restricted by geometry to a line, there is no dynamical degrees of freedom associated with them. Hence the arrangement of fluxes can be purely determined by the charge configuration, meaning that the Hamiltonian can be expressed purely in terms of the charge degrees of freedom, at the cost of introducing non-local terms. This is not necessarily a problem for DMRG, since the MPO representation still has a quite small bond dimension. To construct the Hamiltonian, this approach relies on counting the number of particles to the left (or right) of a given site on the lattice, and have hitherto only been implemented for finite-size systems with open boundary conditions (OBC), where the lattice boundary provides a zero point for counting the particle number. The purpose of this paper is to show that this method can also be applied to infinite Matrix Product States (iMPS), using a carefully constructed MPO that incorporates the correct boundary terms to implement the topological $\theta$ term. We demonstrate this using the Schwinger model.

\section{Construction of the MPO}
\subsection{Lattice Mapping}
In the temporal gauge, $A_0 = 0$, the continuum Schwinger Hamiltonian can be written as:
\begin{align}
    \mathcal{H} &= \int dx\, \left( i \bar\psi (ig \gamma^1 A_1 -\gamma^1 \partial_1) \psi\right) \nonumber \\&\quad + \int dx\,\left(m \bar \psi \psi + \frac{1}{2} (\dot A_1)^2\right).
\end{align}

To represent this Hamiltonian on a lattice, we employ a Kogut-Susskind discretisation \cite{staggeredFermions, kogut_hamiltonian_1975}. This maps the Hamiltonian onto a lattice of $N$ sites with spacing $a$. This approach to discretising the Schwinger model has been used extensively in finite lattice studies \cite{byrnes_density_2002, byrnes_density_2003, szyniszewski_numerical_2013, banuls_mass_2013, banuls_density_2017, buyens_finite-representation_2017, buyens_real-time_2017}.

Assigning a two-component spinor on each site of the lattice is well known to result in fermionic doubling in the continuum limit, $N\rightarrow\infty$ \cite{susskind_lattice_1977,bodwin_lattice_1987,zohar_formulation_2015}. To circumvent this problem, Kogut and Susskind utilise the staggered fermion formulation, wherein the fermionic field is decomposed such that even-numbered lattice sites correspond to particle states and odd-numbered sites correspond to antiparticle states \cite{staggeredFermions}. In Kogut and Susskind's approach, the bosonic degrees of freedom are represented by the gauge links connecting adjacent lattice sites \cite{staggeredFermions, kogut_hamiltonian_1975, kogut_three_1978}. Consequently, the discretised Hamiltonian is expressed as:
\begin{align}
    \nonumber H &=  \frac{-i}{2a} \sum_{n=0}^{N-2} \left(\phi^\dagger_n e^{i\theta_n} \phi_{n+1} - \text{h.c}\right)
    \\
    &\quad+m \sum_n^{N-1}\left((-1)^n \phi^\dagger_n \phi_n \right) + \frac{ag^2}{2} \sum_{n=0}^{N-1} L_n^2 \; ,
\end{align}
where $\theta_n$ is the gauge variable on the link between sites $n$ and $n+1$, and $L_n$ is the conjugate variable such that $[\theta_n, L_m]=i\delta_{nm}$.

Commonly, the gauge degrees of freedom are integrated out so that we are left only with fermionic degrees of freedom
\cite{banuls_mass_2013, banuls_density_2017, banuls_review_2020}. This is done by recognising that the confinement of flux stemming from Gauss' Law in 1+1D allows us to write the bosonic contribution to the Hamiltonian at a site, $n$, purely as a function of the fermionic site occupations for all sites $\leq n$. This is because in one spatial dimension, the individual outgoing flux contribution from a site $n$, $\hat{q}_n$, can only travel along the chain, and thus is summed with all prior incoming flux contributions from previous sites. To this end, we apply Gauss' law, which on the lattice reads \cite{banuls_mass_2013}:
\begin{align}
    L_n - L_{n-1} = \phi^\dagger_n \phi_n - \frac{1}{2}(1-(-1)^n) \; .
\end{align}
This allows us to write $L_n$ as
\begin{align}
    \label{L_n}
    L_n = \ell + \sum_{k=0}^n \left(\phi^\dagger_k\phi_k - \frac{1}{2} \left(1-(-1)^k\right)\right),
\end{align}
where $\ell$ represents the strength of the background field. Departing slightly from the typical notation, we can write this in terms of the \emph{charge operator} at the $k^\text{th}$ site, $q_k$, which is $0$ in the vacuum state, and $\pm 1$ for particle states and antiparticle states respectively. This already takes into account the chiral anomaly, whereby the vacuum state has a shifted particle number, due to all antiparticle states being occupied. (citation here, perhaps Coleman?). These operators are clarified in table \ref{tab:basis-states}, where we also introduce the fermion number operator $n^{(f)} = (1 + (-1)^q)/2$, which counts the number of physical fermions; $+1$ for particle or antiparticle states, or 0 for the quantum vacuum. Charge conservation requires that the overall electric charge $Q = \sum_n q_n = 0$, which is a conserved quantity that can be readily imposed in MPS formulations.

\begin{table}[h]
\centering
\caption{Local basis states and quantum numbers for staggered fermions}
\begin{tabular}{llccc}
\toprule
\textbf{Site Type} & \textbf{State} & \( \phi^\dagger_n \phi_n \) & \( q_n \) & \( n^{(f)}_n \) \\
\midrule
Even (fermion)     & Vacuum         & 0 & 0 & 0 \\
Even (fermion)     & Particle       & 1 & \( +1 \) & 1 \\
Odd (antifermion)  & Antiparticle   & 0 & \( -1 \) & 1 \\
Odd (antifermion)  & Vacuum         & 1 & 0 & 0 \\
\bottomrule
\end{tabular}
\label{tab:basis-states}
\end{table}

In terms of these operators, the electric field operator takes the simpler form,
\begin{equation}
  L_n = \ell + \sum_{k=0}^n q_k \; .
\end{equation}
We can also rewrite the mass term in terms of the total fermion number $N^{(f)} = \sum_n n^{(f)}_n$ as,
\begin{equation}
  \sum_{n=0}^{N-1} (-1)^n \phi^\dagger_n \phi^{}_n = -\frac{N}{2} + \sum_{n=0}^{N-1} n^{(f)}_n \; ,
\end{equation}
where the constant $-N/2$ is a trivial energy shift, and we omit it henceforth.

The gauge link in the kinetic term can be removed through local U(1) gauge transformations \cite{hamer_series_1997}:
\begin{align}
    \label{localTransformations}
    \phi_n &\rightarrow \prod_{k=0}^{n-1}\left(e^{-i\theta(k)}\right)\phi_n,\nonumber\\  \phi^\dagger_n &\rightarrow \prod_{k=0}^{n-1}\left(e^{i\theta(k)}\right)\phi^\dagger_n.
\end{align}
Applying these transformations, the kinetic term transforms in a manner which removes the final gauge dependence of the Hamiltonian:
\begin{align}
    \nonumber H_K &= \left(\phi^\dagger_n e^{i\theta(n)} \phi_{n+1}\right)' \\ \nonumber&= \prod_{k=0}^{n-1}\left(e^{i\theta(k)}\right) \phi^\dagger_n e^{i\theta(n)} \prod_{k=0}^{n}\left(e^{-i\theta(k)}\right) \phi_{n+1}\\
    \nonumber &= \prod_{k=0}^{n-1}\left(e^{i\theta(k)}\right) \prod_{k=0}^{n-1}\left(e^{-i\theta(k)}\right) \phi^\dagger_n  \phi_{n+1} \\
    &= \phi^\dagger_n  \phi_{n+1}.
\end{align}

Hence, the Hamiltonian can be written solely in terms of fermionic degrees of freedom as:
\begin{align}
    \label{dimensionHamiltonian}
    \nonumber H &=  \frac{-i}{2a} \sum_{n=0}^{N-2} \left(\phi^\dagger_n \phi_{n+1} - \text{h.c}\right)
    +m \sum_{n=0}^{N-1} n^{(f)}_n \\
    & + \frac{ag^2}{2} \sum_{n=0}^{N-1} \left(\ell+\sum_{k=0}^n q_k \right)^2.
\end{align}

As a consequence of integrating out the gauge degrees of freedom, we introduce non-local interactions in the electric field term. Because of this, the removal of gauge degrees of freedom from the lattice Schwinger Hamiltonian has generally been deemed an option that is only available to algorithms utilising OBCs \cite{banuls_density_2017, papaefstathiou_density_2021}, and those attempting infinite simulations have thus far opted to skip this step \cite{zapp_tensor_2017}, sacrificing the efficiencies it affords.

\subsection{The Free Schwinger Model}
We now turn to constructing Matrix Product Operator representations of the Hamiltonian. Following the standard approach in the literature, we seek not to find an MPO representation of the lattice Hamiltonian, but of a dimensionless form of the lattice Hamiltonian, $W = \frac{2}{ag^2}H$ \cite{banuls_mass_2013, banuls_review_2020}.

\begin{figure}[ht]
\centering
\includegraphics[width=0.45\textwidth]{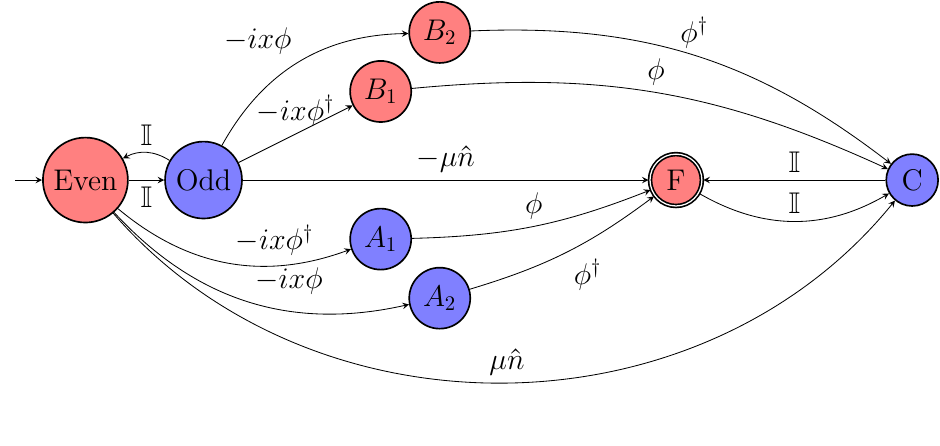}
\caption{Finite state machine representation of operator string states in the free Schwinger Hamiltonian. We represent the two-site nature of the MPO unit cell through a bipartite graph, where a state corresponding to an existing string of even length (red) can only travel to a state corresponding to an existing string of odd length (blue) and vice-versa.}
\label{fig:freeSchwinger_fsm}
\end{figure}

We first approach the MPO for the dimensionless free Schwinger model, where the electric field term is set to 0, rendering the Hamiltonian invariant under the removal of the gauge degrees of freedom. This variant of the model therefore has no non-locality, and hence can illustrate precisely how the iDMRG algorithm should work in theory. Its Hamiltonian is given by
\begin{align}
    \label{freeSchwinger}
    H_{\text{free}} &= -ix \sum_{n=0}^{N-2}\left(\phi^\dagger_n \phi_{n+1} + \phi_n \phi^\dagger_{n+1} \right) \nonumber\\ &\quad\quad + \mu \sum_{n=0}^{N-1} n_n^{(f)},
\end{align}
where we have used the lattice parameters $x=\frac{1}{g^2a^2}$ and $\mu = \frac{2m}{g^2a}$ to convert our Hamiltonian to its dimensionless counterpart. The kinetic term can be written as a string of operators. For example,
\begin{align}
    \nonumber \sum_{n=0}^{N-2} \phi^\dagger_n \phi_{n+1} &=\phi^\dagger_0 \otimes \phi_1 \otimes\mathbb{I}\otimes\mathbb{I}\otimes\mathbb{I}\otimes \ldots \\
    \nonumber &+\mathbb{I}\otimes \phi^\dagger_1 \otimes \phi_2 \otimes\mathbb{I}\otimes\mathbb{I}\otimes \ldots \\
    \nonumber &+\mathbb{I}\otimes\mathbb{I}\otimes \phi^\dagger_2 \otimes \phi_3 \otimes\mathbb{I}\otimes \ldots \\
    \nonumber &\quad\vdots\\
     &+\mathbb{I} \otimes \ldots \otimes \phi^\dagger_{N-2} \otimes \phi_{N-1}.
    \label{freeSchwinger_tensorProd}
\end{align}
The mass term is similarly written, with the $\mu$ parameter omitted, as
\begin{align}
    \label{massTensorProdFreeSchwinger}
    \sum_{n=0}^{N-1} n_n^{(f)} &= \sum_{n=0}^{N-1}\mathbb{I}^{\otimes n} \otimes n_n^{(f)} \otimes \mathbb{I}^{\otimes\left(N-n-1\right)}.
\end{align}
Using these strings, the free model's translationally invariant MPO can be written as a two-site unit cell, represented by the FSM in Figure \ref{fig:freeSchwinger_fsm} and the operators $W_0$ and $W_1$, defined as
\begin{align}
    W_j = \left( \begin{matrix}
   \mathbb{I}& -ix\phi & -ix\phi^\dagger & (-1)^j  \mu\phi^{\dagger}\phi \\
    0 & 0 & 0 & \phi^\dagger \\
    0 & 0 & 0 & \phi \\
    0 & 0 & 0 & \mathbb{I}
    \end{matrix} \right),
    \label{freeSchwingerMPO}
\end{align}
where $j \in \{0,1\}$.

\subsection{The Interacting Model}
\label{sec:InteractingModel}
Unlike the free model, the construction of an MPO to represent the interacting model is non-trivial. This is because the electric field term for a given site is given as a function of all previous site occupancies. This is problematic in the context of an infinite lattice, as there is no boundary from which we can ''begin" counting. We sought to rectify this by encoding the off-site information into the \textit{effective Hamiltonian block operators} which contract with our MPOs in the iDMRG process.

Typically, we can represent the contracted block of sites to the left of a given unit cell using the $E$ and $F$ effective Hamiltonian block operators respectively. These are vectors of operator matrices which, when contracted with the MPO of that unit cell, $W$, yield the Hamiltonian. That is, $E(n)$ ($F(n)$) represents the effective Hamiltonian up to site $n$ counting up from left to right (from right to left), and the contraction, $E(n)\circ W^{[n+1]} \circ F(n)$ recovers the full Hamiltonian of the system \cite{schollwock_density-matrix_2005, mcculloch_infinite_2008}. For an upper-triangular MPO, the $E$ and $F$ vectors can be written as \cite{mcculloch_infinite_2008, phien_infinite_2012, michel_schur_2010}
\begin{align}
    B_\alpha (n+1) &= T_{W_{\alpha \alpha}^{[n+1]}}(B_\alpha (n)) \nonumber\\&\quad + \sum_{\beta < \alpha} T_{W_{\beta \alpha}^{[n+1]}} (B_\beta (n)), \label{EFBlock_calcs}
\end{align}
where $B \in \{E, F\}$ and the transfer operator, $T_X(B)$, is defined as
\begin{align}
    T_X(B) = \sum_{ss'} \langle s  | X | s'\rangle A^{s' \dagger} B A^s.
\end{align}
Because of the recursive nature of Equation \eqref{EFBlock_calcs}, these block operators are usually defined with boundary conditions
\begin{align}
    E_\alpha(0) = \delta_{\alpha,0} \mathbb{I}\nonumber\\
    F_\beta(0) = \delta_{\beta,d-1} \mathbb{I},
    \label{blockOperatorBoundaries}
\end{align}
where $d$ is the dimension of the MPO. This condition effectively represents the assumption that there is nothing beyond the left and right of these sites. Where interactions are restricted to being between sites with finite separation, the error from this assumption reduces to zero as the lattice size grows to the bulk limit. However, because of the non-local interaction present in Equation \eqref{dimensionHamiltonian}'s final term, the error does not reduce as we increase the number of sites in our lattice, so long as our wavefunction is pre-existing or is in the presence of a background electric field.

To address this, we intervened in the iDMRG algorithm to modify $E(0)$ and $F(0)$ to instead represent the electric field contribution from outside the boundaries. This involved calculating the elements, $E_\alpha(n)$ and $F_\alpha(n)$ $\forall \alpha \in \{0,d-1\}$, for arbitrary $n$ and background field, $\ell$.

Because of the structure of \eqref{dimensionHamiltonian}'s electric field term, the $E(n)$ block's "corrections" could all be expressed as a function of the charge to the left of site $n$. Under the U(1) symmetry of our system, this charge is a conserved quantum number. Hence, working in the basis of this quantum number, we could construct and apply a charge operator as part of the $E(n)$ corrections. However, the electric field term as it appears in \eqref{dimensionHamiltonian} is asymmetric, and as a result, we found that some of the corrections required for the $F$ block diverged. Consequently, these corrections could not be expressed as a function of the charge operator or any other local operator.

To rectify this we sought to express the electric field term in a symmetric form, which gives the electric field as a function of all sites across both sides of the lattice, rather than just the left side. To this end, we can write the electric field at a site, $n$, as a function of the background field, $\ell$, the charge stemming from ''beyond" the left boundary, which we denote $b_L$, and the charge stemming from all site occupations within the boundaries of the lattice:
\begin{align}
    L_n = \ell + b_L + \sum_i^n \hat{q}_i.
\end{align}

Exploiting the charge conservation of the Schwinger model, we can alternatively express the electric field at site $n$ as a function of the charge to the \textit{right}:
\begin{align}
    L_n = \ell -\sum_{j = n+1}^{N-1} \hat{q}_j - b_R,
\end{align}
where $b_R$ represents the boundary charge on the right. As our lattice grows, charge conservation constrains $\sum_i^n \hat{q}_i \overset{!}{=} - \sum_{j = n+1}^N \hat{q}_j$. Hence, to maintain charge conservation, we must also require that our boundary corrections are such that $b_L = -b_R$. Hence, we can rewrite $L_n^2$ as
\begin{align}
   \sum_{n} L^2_n &= \sum_{n}  \left(\left( \ell  + b_L + \sum_{i}^n \hat{q}_i \right)\left(\ell -\sum_{i=n+1}^{N-1} \hat{q}_i - b_R\right)\right) \nonumber \\
    &= \sum_{n}
    \left((\ell +b_L)^2 + 2 (\ell + b_L)\sum_{i}^n \hat{q}_i \right)
    \nonumber\\ &\quad - \sum_{j>i} \left(\left(j-i\right) \hat{q}_i \hat{q}_j \right),
    \label{H_E_symDer}
\end{align}
where we use the fact that
\begin{align}
    \sum_{n=0}^N \left(
    \sum_{i=0}^n \hat{q}_i \sum_{j=i+1}^N \hat{q}_j
    \right) = \sum_{j>i}\left((j-i)\hat{q}_i \hat{q}_j \right). \label{SumOverSymmetricEquiv}
\end{align}
While the terms in Equation \eqref{H_E_symDer} involving the boundary corrections and background field are not symmetric due to insignificant simplifications, the important modification here is the symmetry in the final term, which is repeated in Equation \eqref{SumOverSymmetricEquiv}. This term directly illustrates the confining nature of the Schwinger model. This stems from the prefactor, $(j-i)$, which shows that the energy contribution from the potential between sites $i$ and $j$ grows linearly with the distance between $i$ and $j$. This term can be expressed, up to a sign change, using the single-site MPO,
\begin{align}
    W_E = \left(
        \begin{matrix}
            \mathbb{I} & \frac{1}{\sqrt{2}}\hat{q} & \frac{1}{\sqrt{2}}\hat{q} & 0 \\
            0 & \mathbb{I} & 2\mathbb{I} & \frac{1}{\sqrt{2}}\hat{q}\\
            0 & 0 & \mathbb{I} & \frac{1}{\sqrt{2}}\hat{q} \\
            0 & 0 & 0 & \mathbb{I}
        \end{matrix}
    \right),
    \label{singleSiteHE}
\end{align}

\begin{figure}[ht]
\centering
\includegraphics[width=0.45\textwidth]{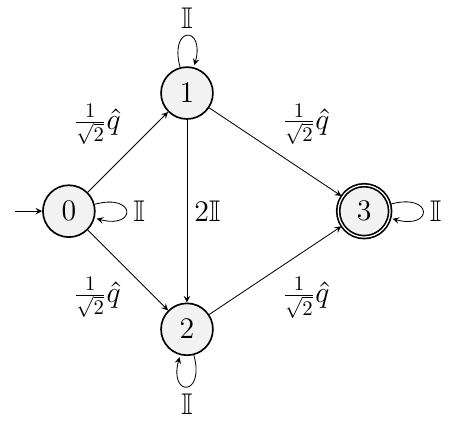}
\caption{Finite state machine representation of the confining term of the interacting Schwinger Hamiltonian.}
\label{fig:confining_fsm}
\end{figure}

Here, the linear increase in the confining potential stems from the loop transitions on states $1$ and $2$, which correspond to the diagonal elements, $(W_E)_{11}$ and $(W_E)_{22}$, as well as the $2\mathbb{I}$ transition from state $1$ to $2$, which corresponds to $(W_E)_{12}$. The number of unique paths of length $n$ starting at state $1$ and ending at state $2$ is equal to $n$. For example, the unique paths for $n\in [1,3]$ are given below:
\small
\begin{align}
\begin{matrix}
        n=1\quad & n=2\quad & n=3\quad \\
        1\rightarrow 2\quad &
    \begin{matrix}
    1 \circlearrowright 1 \rightarrow 2 \\ 1 \rightarrow 2 \circlearrowright 2
    \end{matrix}\quad &
    \begin{matrix}
    1 \circlearrowright 1 \circlearrowright 1 \rightarrow 2 \\ 1 \circlearrowright 1 \rightarrow 2 \circlearrowright 2 \\   1  \rightarrow 2 \circlearrowright 2  \circlearrowright 2
    \end{matrix}\quad
\end{matrix}
\end{align}
\normalsize
Each unique path can be characterised by when the $1\rightarrow 2$ transition occurs. This transition can occur at any point in the path, but because it is unidirectional, it can only occur once in the path. Hence, the number of unique paths of length $n$ is equal to $n$. Any path that includes the $1\rightarrow 2$ transition will necessarily include a $0\rightarrow 1$ and $2 \rightarrow 3$ transition. Such a path which picks up $q_i$ in the $0\rightarrow 1$ transition and $q_j$ in the $2\rightarrow 3$ transition will be of length $j-i+1$, with a sub-length from state $1$ to state $2$ of $j-i-1$. There will therefore be $j-i-1$ unique paths which include a $1\rightarrow 2$ transition and pick up $q_i$ and $q_j$. This set of paths therefore contributes
\begin{align}
    \frac{1}{\sqrt{2}}\frac{1}{\sqrt{2}} 2(j-i-1) \hat{q_i} \hat{q_j}  = \sum_{j > i} (j-i-1) \hat{q_i} \hat{q_j} \label{1to2MPO}
\end{align}
to our confining term, where kronecker products of the Identity are left implicit, the factor of $\frac{1}{\sqrt{2}}\frac{1}{\sqrt{2}}$ comes from the $0\rightarrow 1$ and $2\rightarrow 3$ transitions, and the factor of $2$ is picked up from each $1\rightarrow 2$ transition. Finally, for any pair of sites, $i$ and $j$ separated by path length $n$, there are only two unique paths which do not involve the $1\rightarrow 2$ transition. For example, the unique paths for $n \in [2,4]$ are:
\small
\begin{align}
\begin{matrix}
        n=2\quad & n=3\quad & n=4 \quad\\
        \begin{matrix}
        0 \rightarrow 2 \rightarrow 3 \quad  \\
        0 \rightarrow 1 \rightarrow 3 \quad
        \end{matrix} &
    \begin{matrix}
    0 \rightarrow 2 \circlearrowright 2 \rightarrow 3 \\
    0 \rightarrow 1 \circlearrowright 1 \rightarrow 3
    \end{matrix}\quad &
    \begin{matrix}
    0 \rightarrow 2 \circlearrowright 2 \circlearrowright 2 \rightarrow 3  \\
    0 \rightarrow 1 \circlearrowright 1 \circlearrowright 1 \rightarrow 3
    \end{matrix}
\end{matrix}
\end{align}
\normalsize
These paths therefore contribute
\begin{align}
   \sum_{j > i} \frac{1}{\sqrt{2}}\frac{1}{\sqrt{2}} 2 \hat{q_i} \hat{q_j}  = \hat{q_i} \hat{q_j} \label{not1to2MPO}
\end{align}
to our confining term. Summing up Equations \eqref{1to2MPO} and \eqref{not1to2MPO}, we see that $W_E$ yields our confining term, Equation \eqref{SumOverSymmetricEquiv}:
\begin{align}
   \sum_{j > i} (j-i-1) \hat{q_i} \hat{q_j} + \sum_{j > i}\hat{q_i} \hat{q_j} = \sum_{j > i}(j-i) \hat{q_i} \hat{q_j}
\end{align}

We can add the remaining terms and correct the sign error by using the below boundary vectors:
\begin{align}
    L = \left(
    \begin{matrix}
        \frac{\ell + b_L}{\sqrt{2}} & \frac{(\ell + b_L)^2}{2} & 1 & \frac{\ell + b_L}{\sqrt{2}}
    \end{matrix}
    \right), \\ R = \left(
    \begin{matrix}
        0 & 0 & 1 & \frac{-\sqrt{2}}{\ell + b_L}
    \end{matrix}\right)^T.
\end{align}

To verify that these boundary vectors produce the required Hamiltonian term, we can calculate
\begin{align}
    L \circ \left( \bigcirc_{n=0}^{N-1} W_E \right) \circ R.
\end{align}
Calculating $\bigcirc_{n=0}^{N-1} W_E$ via an inductive proof gives
\begin{widetext}
\begin{align}
    \bigcirc_{n=0}^{N-1} W_E = \left(
        \begin{matrix}
            \mathbb{I} & 0 & -\frac{1}{\sqrt{2}} \sum_{n=0}^{N-1} 2n \hat{q_n} & \sum_{j>i}^{N-1} (j-i) q_i q_j \\
            0 & \mathbb{I} & 2N\mathbb{I} & \frac{1}{\sqrt{2}} \sum_{n=0}^{N-1} 2n \hat{q_n}\\
            0 & 0 & \mathbb{I} & 0 \\
            0 & 0 & 0 & \mathbb{I}
        \end{matrix}
    \right),
    \label{bigcircN}
\end{align}
where we have once again used the constraint, $\sum_{n=0}^{N-1} \hat{q_n} = 0$. Hence, our Hamiltonian term is
\begin{align}
    H_E &= L \circ \left( \bigcirc_n^{N-1} W_E \right) \circ R \nonumber \\& = -2(\ell + b_L) \sum_{n=0}^{N-1} n \hat{q}_n + N(\ell + b_L)^2 - \sum_{j>i}^{N-1} (j-i) \hat{q}_i \hat{q}_j \nonumber\\
    &= \sum_{n=0}^{N-1} \left((\ell+b_L)((\ell+b_L) + 2\sum_{i=0}^n \hat{q}_i) \right) - \sum_{j>i}^{N-1} (j-i) q_i q_j \nonumber\\
    &= \sum_{n=0}^{N-1} L_n^2,
\end{align}
as required, where we have again used the conservation of charge, allowing
\begin{align}
    \sum_{n=0}^{N-1} \sum_{i=0}^n \hat{q}_i = \sum_{n=0}^{N-1} (N-n) \hat{q}_n = -\sum_{n=0}^{N-1} n \hat{q}_n.
\end{align}
\end{widetext}

In addition to representing the linearly increasing confining potential between particles in a more intuitive manner than the initial representation of the Hamiltonian, this representation also yields boundary corrections which can be expressed as a function of local operators. Constructing the Schwinger model on an infinite lattice in such a way has several advantages.

\subsection{Generalisation}
The linearly increasing potential between two charges according to their separation is not a property unique to the Schwinger model, and hence, the methods here can be generalised to other models with similar characteristics.

\begin{figure}[h]
\centering
\includegraphics[width=0.45\textwidth]{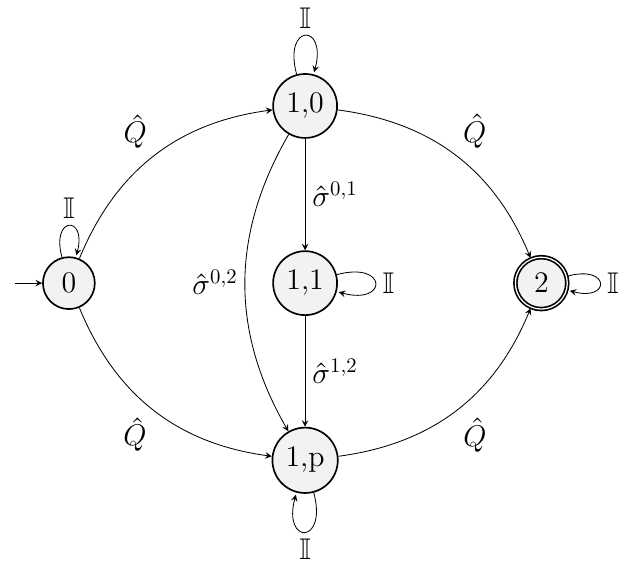}
\caption{Finite state machine representation of a quadratic confining potential.}
\label{fig:confining_fsm_quadratic}
\end{figure}

To begin, we generalise Equation \eqref{singleSiteHE}'s MPO and the corresponding Figure \ref{fig:confining_fsm}. For example, we can generalise the linear strength of the confining potential to higher-degree strengths by adding additional intermediary states between states $1$ and $2$ in Figure \ref{fig:confining_fsm}'s FSM, as in Figure \ref{fig:confining_fsm_quadratic}. This FSM can be written as a $5\times5$ MPO:
\begin{align}
    W = \left(
        \begin{matrix}
            \mathbb{I} & \hat{Q} & 0 &\hat{Q} & 0 \\
            0 & \mathbb{I} & \hat \sigma_0^1 & \hat \sigma_0^2  & \hat{Q}\\
            0 & 0 & \mathbb{I} & \hat \sigma_1^2 & 0\\
            0 & 0 & 0 & \mathbb{I} & \hat{Q} \\
            0 & 0 & 0 & 0 & \mathbb{I}
        \end{matrix}
    \right).
    \label{quadraticMPO}
\end{align}

Excluding identities trailing infinitely to the left and right, two charges separated by a distance $(j-i)$ and described under this MPO yield the operator string sum,
\begin{align}
    H_{i,j} = &2\left(\hat{Q_i} \otimes \mathbb{I}^{\otimes j-i-1} \otimes \hat{Q_j}\right) \nonumber \\
    +&  \sum_{\alpha, \beta, \gamma} \hat{Q_i} \otimes \mathbb{I}^{\otimes \alpha } \otimes  \hat{\sigma}^{0,1}\mathbb{I} \otimes \mathbb{I} ^{\otimes \beta } \otimes \hat{\sigma}^{1,2} \mathbb{I} \otimes \mathbb{I}^{\otimes \gamma} \otimes \hat{Q_j} \nonumber \\
   +& \sum_{\delta, \epsilon} \hat{Q_i} \otimes \mathbb{I}^{\otimes \delta } \otimes  \hat{\sigma}^{0,2}\mathbb{I} \otimes \mathbb{I} ^{\epsilon} \otimes \hat{Q_j}.
\end{align}
where we replace $\frac{1}{\sqrt{2}}\hat{q}$ with the more general $\hat{Q}=c\hat{q}$, $c$ is some constant scalar factor, and we apply the constraints,
\begin{align}
    \alpha + \beta + \gamma + 3 = \delta + \epsilon + 2 = j-i.
\end{align}
Defining $\hat\sigma_a^b = \{\sigma_a^b \mathbb{I} : \sigma_a^b \in \mathbb{C}\}$ and omitting local identity operators, we get
\begin{align}
    H_{i,j} = c^2 \left(2 + \sum_{\alpha, \beta, \gamma}\sigma_0^1\sigma_1^2 + \sum_{\delta, \epsilon} \sigma_0^2 \right) \hat q_i \hat q_j.
\end{align}
Using our constraints and defining $r$ as the separation between two sites, $r=j-i$, we can apply the stars-and-bars theorem to restate this independent of $\alpha, \beta, \gamma, \delta, \epsilon$:
\begin{align}
      H_{i,j} &= c^2 \left(2 + \binom{r-1}{2} \sigma_0^1\sigma_1^2 + \binom{r-1}{1} \sigma_0^2 \right) \hat q_i \hat q_j.
\end{align}
This yields a potential which grows quadratically with $r$.

\begin{figure}[ht]
\centering
\includegraphics[width=0.45\textwidth]{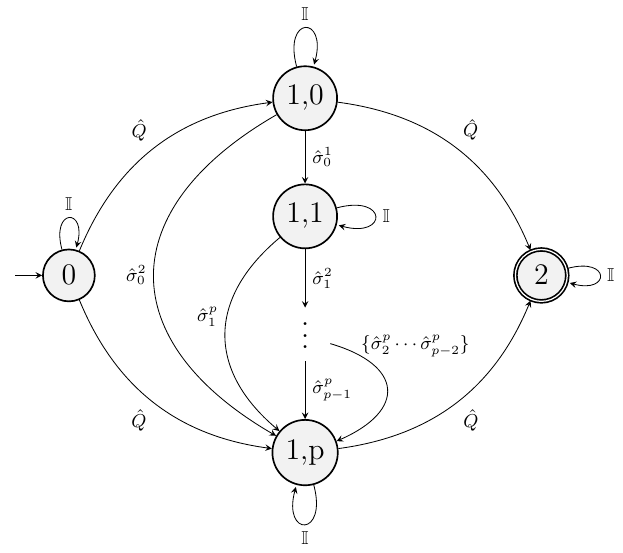}
\caption{Finite state machine representation of a confining potential which grows with degree $p$ in charge separation.}
\label{fig:confining_fsm_poly}
\end{figure}

More generally, we find that an FSM of the form given in Figure \ref{fig:confining_fsm_quadratic} which includes $p$ intermediary states between state $1,0$ and state $2$ returns a confining potential which grows with degree $p$ in charge separation, $r$. Figure \ref{fig:confining_fsm_poly} represents this generalised FSM and corresponds to an MPO of size $p+3$:
\begin{align}
    W = \left(
        \begin{matrix}
            \mathbb{I} &  \hat{Q} & 0 & 0 & 0 & \cdots &\hat{Q} & 0 \\
            0 & \mathbb{I} & \hat\sigma_0^{1} & 0 & 0 & \cdots&  \hat\sigma_0^p & \hat{Q}\\
            0 & 0 & \mathbb{I} & \hat\sigma_{1}^{2} & 0 & \cdots & \hat\sigma_1^p & 0\\
            \vdots & \vdots & \ddots & \ddots & \ddots & \ddots & \vdots & \vdots\\
            0 & 0 & 0 & \ddots & \mathbb{I} & \hat\sigma_{p-2}^{p-1}  &  \hat\sigma_{p-2}^{p}  & 0 \\
            0 & 0 & 0 & \cdots & 0 & \mathbb{I} & \hat\sigma_{p-1}^{p} & 0 \\
            0 & 0 & 0 & \cdots &0 &0 & \mathbb{I} & \hat{Q} \\
            0 & 0 & 0 & \cdots &0 &0 & 0 &\mathbb{I}
        \end{matrix}
    \right),
    \label{polynomialMPO}
\end{align}

The strong kronecker product series contracting this MPO with itself, $\left( \bigcirc_{n=0}^{N-1} W \right)$, can be written explicitly as
\footnotesize
\begin{widetext}
\begin{align}
  \begingroup
    \setlength{\arraycolsep}{2pt}   
    \let\oldsum\sum
    \renewcommand{\sum}{\oldsum\limits}
    \let\oldprod\prod
    \renewcommand{\prod}{\oldprod\limits}
     \mathop{\bigcirc}\limits_{n=0}^{\bar N} W =\left(
        \begin{matrix}
            \mathbb{I} &  \varsigma_0 & \pihat{0}{1} \varsigma_1  & \pihat{0}{2} \varsigma_2 &  \cdots &
            \sum_{k=0}^{\bar{N}} \hat{Q}_k \Bigl(1 + \sum_{m=1}^{p} \binom{\bar{N}-k}{m} \pihat{0}{m-1} \hat\sigma_{m-1}^{p} \Bigr)
            & \sum_{k=1}^{\bar{N}}
            \sum_{l=0}^{k-1}
            \left(
            \begin{array}{c}
             2 \hat{Q}_{k} \hat{Q}_{l} \\
             + \sum_{s=1}^{p}
                \hat{Q}_{k}\,\hat{Q}_{l} \\
                \times \binom{k-l-1}{s} \pihat{0}{s-1}  \hat{\sigma}_{s-1}^{p}
            \end{array}
                            \right) \\
            0 & \mathbb{I} & \binom{N}{1}\pihat{0}{1} & \binom{N}{2} \pihat{0}{2}  &\cdots&
            \sum_{k=0}^{p-1} \binom{N}{k+1} \pihat{0}{k} \hat\sigma_{k}^{p}
            & \sum_{k=0}^{\bar{N}} \hat{Q}_k \left(1 + \sum_{m=1}^p \binom{k}{m} \pihat{0}{m-1}  \hat\sigma_{m-1}^{p}\right)  \\
            0 & 0 & \mathbb{I} & \binom{N}{1}\pihat{1}{2} &  \cdots &  \sum_{k=0}^{p-2} \binom{N}{k+1} \pihat{1}{k+1} \hat\sigma_{k+1}^{p} & \sum_{k=1}^{\bar{N}} \hat{Q}_k \sum_{m=1}^{p-1} \binom{k}{m} \pihat{1}{m}  \hat\sigma_{m}^{p}
            \\
            \vdots & \vdots & \ddots & \ddots & \ddots  & \vdots & \vdots\\
            0 & 0 & 0 &\cdots & \mathbb{I} & \sum_{k=0}^{p-p} \binom{N}{k+1} \pihat{p-1}{k+p-1} \hat\sigma_{k+p-1}^{p} & \sum_{k=1}^{\bar{N}} \hat{Q}_k \sum_{m=1}^{1} \binom{k}{m} \pihat{p-2}{m+p-2}  \hat\sigma_{m+p-2}^{p}
            \\
            0 & 0 & 0 & \cdots & 0 &\mathbb{I} & \sum_{k=0}^{\bar{N}}\hat{Q}_k  \\
            0 & 0 & 0 & \cdots  &0 & 0 &\mathbb{I}
        \end{matrix}
    \right).
    \label{polynomialMPO_Nn}
    \endgroup
\end{align}

\end{widetext}
\normalsize
where we define $\bar{N}:=N-1$, $\pihat{a}{b} := \prod_{i=a}^{b-1} \sigma_{i}^{i+1}$, and $\varsigma_k := \sum_{j=0}^{\bar{N}} \binom{\bar{N}-j}{k} \hat{Q}_j$,  for typesetting reasons. Note also that our family of FSMs do not admit any transitions with labels $\hat \sigma_{k}^l$ where $k \geq l$.  Hence, in our family of FSMs, $\hat \sigma_k^l = 0, \; \forall k \geq l$. A proof of this is provided by induction in Appendix \ref{appendix:FSMProof}.

Under application of the typical boundary vectors given by Equation \eqref{blockOperatorBoundaries}, our  generalised confining term for $N$ sites is evaluated as
\begin{align}
    H &= \left( \begin{matrix}
        \mathbb{I} & 0 & \cdots 0
    \end{matrix}\right) \circ \left(\mathop{\bigcirc}\limits_{n=0}^{\bar N} W^{[n]}\right) \circ \left( \begin{matrix}
        0 & 0 & \cdots \mathbb{I}
    \end{matrix}\right)^T \nonumber \\
    &= \sum_{j=1}^{\bar{N}}
            \sum_{i=0}^{j-1} \hat{Q}_{j} \hat{Q}_{i}
            \left(
             2               + \sum_{s=1}^{p}
                       \binom{r-1}{s} \pihat{0}{s-1}  \hat{\sigma}_{s-1}^{p}                 \right),
\end{align}
where $r = j-i$ again represents the separation between two sites. We previously treated $\hat{Q}_k$ as the charge for site $k$ multiplied by some constant, $c$. However, by reference to Figure \ref{fig:confining_fsm_quadratic}, it is trivial to see that, if we set the first $\hat{Q}$ transitions of a given path as $\hat{q} \mu$, and the second $\hat{Q}$ transitions of a given path as $\hat{q}\nu$, then we can introduce an additional degree of freedom into our confining strength. That is, the confining strength between any two sites, $j,k$, can be written as
\begin{align}
    H_{i,j} = \hat{q}_j \hat{q}_i \mu \nu \left(
             2               + \sum_{s=1}^{p}
                       \binom{r-1}{s} \pihat{0}{s-1}  \hat{\sigma}_{s-1}^{p}                 \right).
\end{align}

To recover a polynomial of order $p$ in $r$, we first choose $\hat\sigma_{a}^{a+1} = \mathbb{I}$ for all $ 0 \leq a \leq p-2$, which gives us
\begin{align}
        H_{i,j} = \hat{q}_j \hat{q}_i \mu \nu \left(
             2               + \sum_{s=1}^{p}
                       \binom{r-1}{s}  \hat{\sigma}_{s-1}^{p}                 \right).
                       \label{polynomialProducingH}
\end{align}
Any polynomial of degree $p$ in $r$ can be expressed as a linear combination of binomial coefficients, $\sum_{s=0}^p c_s \binom{r}{s}$. Equation \eqref{polynomialProducingH} is such a linear combination multiplied by $\hat{q_i}\hat{q_j}$, with the shift of $r \rightarrow r-1$, and setting $ \mu \nu\sigma_{s-1}^{p} = c_s$, and $ \frac{c_0}{\mu \nu} = 2$. For a polynomial of degree $p$ in $r$, represented as a linear combination of binomial coefficients, $f(r) = \sum_{s=0}^p c_s \binom{r}{s}$, the coefficients $c_s$ are known to be equal to
\begin{align}
    c_s = \sum_{i=0}^s (-1)^{s-i} \binom{s}{i} f(i).
\end{align}
However, because $H_{i,j}$ represents a linear combination which shifts $r \rightarrow r-1$, we instead must recover the coefficients using $f(r) = \sum_{s=0}^p c_s \binom{r-1}{s}$, or, equivalently:
\begin{align}
    f(r+1) = \sum_{s=0}^p c_s \binom{r}{s}.
\end{align}
Hence, we recover our coefficients as
\begin{align}
    c_s = \sum_{i=0}^s (-1)^{s-i} \binom{s}{i} f(i+1).
\end{align}

We can generalise further by altering Figure \ref{fig:confining_fsm_poly}'s FSM to allow for non-identity loops on $1,0$ to $1,p$. It can be trivially verified that this permits confining potentials which grow exponentially with separation.


\section{Model Verification}
In this section, we briefly verify that our model reproduces the expected features of the Schwinger model.

\begin{figure}[t]
  \centering
  \resizebox{\columnwidth}{!}{\input{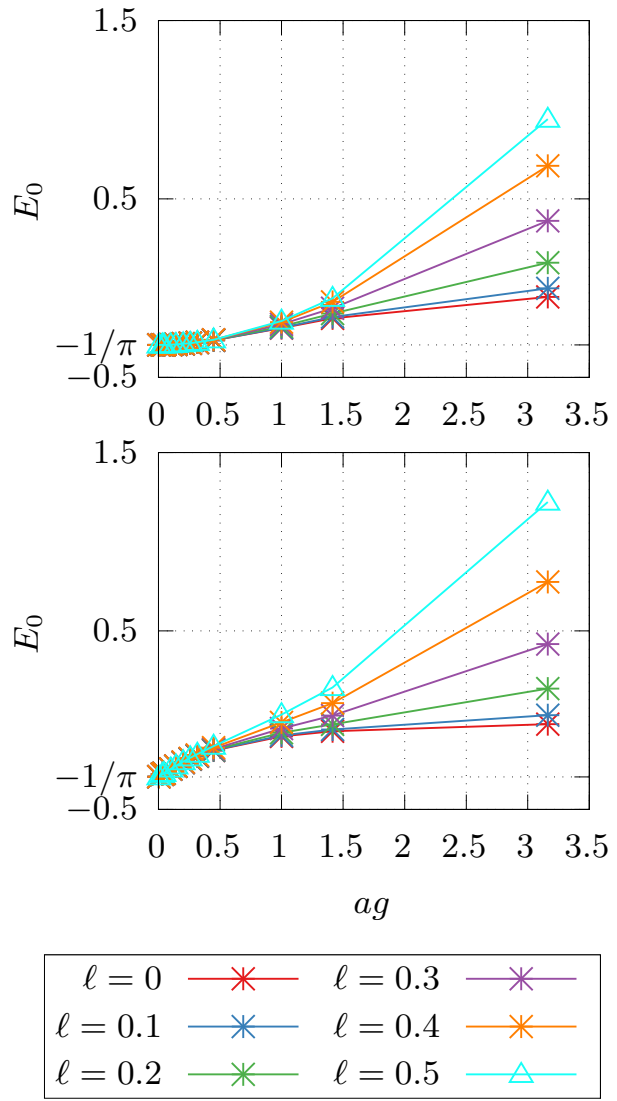}}
  \vspace{-7em}
  \caption{Ground-state energies for $\ell\in\{0.0,0.1,\dots,0.5\}$ versus $ag$ for $m/g=0$ (top) and $m/g=1$ (bottom).}
  \label{gnu:indx}
\end{figure}

In the continuum limit, $x\rightarrow\infty$, the hopping term Hamiltonian dominates, and the Schwinger model becomes equivalent to the XY model \cite{crewther_eigenvalues_1980}, whose ground state energy is exactly known to be $\frac{-1}{\pi}$ \cite{mattis_history_1981}. Hence, in these limits, we expect our model to produce this ground state energy, independent of mass or background field \cite{crewther_eigenvalues_1980, szyniszewski_numerical_2013}. To set $x\rightarrow\infty$, we take $ag\rightarrow 0$, and indeed, in Figure \ref{gnu:indx}, we see ground state energies tend to $-1/\pi$ in this limit.

Of course, in the continuum limit, the interaction term of our Hamiltonian vanishes, as $ag\rightarrow0$, and hence Figure \ref{gnu:indx} only verifies our model insofar as it represents the \textit{free} Schwinger model. Accordingly, we now identify characteristics relating to the critical point of the interacting model, $\left(\frac{m}{g}\right)_c$, which is known to appear at $\ell = 0.5$ (analytically), and $\left(\frac{m}{g}\right)_c \approx 0.334$ (numerically) \cite{byrnes_density_2002, byrnes_density_2003, Fujii2025Critical}.

We begin by calculating the string tension, defined as the difference between vacuum energies with and without a background field per unit length, $a$ \cite{hamer_massive_1982, szyniszewski_numerical_2013}:
\begin{align}
    T(\theta) = \frac{E_0(\theta) - E_0(0)}{a}.
    \label{stringTension_1}
\end{align}
For very small masses, $T(\theta)$ is known to be \cite{coleman_more_1976, hamer_massive_1982, byrnes_density_2002}
\begin{align}
    T(\theta) = \frac{cmg}{\sqrt{\pi}} (1-\cos(\theta)),
    \label{lowMassTension}
\end{align}
while in the non-relativistic limit, $m/g\rightarrow \infty$, we expect \cite{hamer_massive_1982, byrnes_density_2002}
\begin{align}
    T(\theta) =\frac{\theta^2}{2}  \frac{g^2}{\left(2\pi \right)^2}.
    \label{nonrelativisticE}
\end{align}
This occurs because the fermion density decreases disproportionately to mass due to the rising energetic cost of particle excitations. Consequently, as the number of excitations approaches zero, the ground state energy becomes proportionate to the square of the background field.

\begin{figure*}[t]
  \centering
  \resizebox{\textwidth}{!}{\input{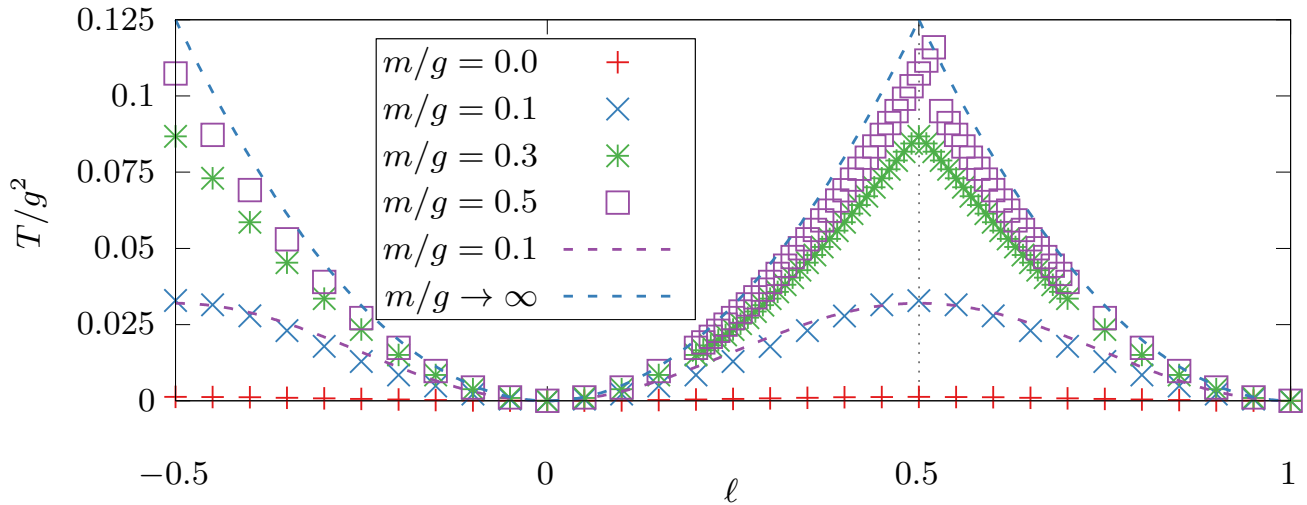}}
  \caption{Dimensionless string tension numerical results (points) plotted alongside analytical predictions (dashed lines) for small mass, $m/g=0.1$, and non-relativistic limit, $m/g\to\infty$ for $x=1000$, $\chi=250$. A vertical dotted line emphasises the critical background field, $\ell=0.5$.}
  \label{gnu:stringTension}
\end{figure*}

At the critical background field, $\theta=k\pi$ (or $\ell=0.5k$) for some integer multiple $k$, the non-relativistic expression yields a discontinuity. In Figure \ref{gnu:stringTension}, we plot the dimensionless string tension ($T/g^2$) numerically and analytically. Our numerical results match expectations. As our mass-to-coupling ratio approaches the critical point, we see the string tension peaks around $\ell=0.5$ getting sharper and approaching the shape of the non-relativistic string tension. This discontinuity in the gradient of our higher mass slopes indicates spontaneous symmetry breaking and the presence of a first-order phase transition above the critical mass, $\left(\frac{m}{g}\right)_c$. At $m/g=0.5$, well above $\left(\frac{m}{g}\right)_c$, we see a clear discontinuity. Interestingly, this discontinuity occurs at some $\ell > 0.5$. This result encapsulates the existence of a two-fold degeneracy. As we can see, the curve resembles some function of the square of the background field even after passing $\ell=0.5$. This indicates that the wavefunction has become trapped in an excited state which is symmetric about $\ell=0$, and is delayed in its escape to the state symmetric about $\ell=1$. The phenomenon of the wavefunction being caught in a local minimum which is not actually a global minimum is not one which is unfamiliar in DMRG, \cite{schollwock_density-matrix_2011} and the two-state degeneracy of the Schwinger model makes it particularly prone to this behavior.

\begin{figure*}[t]
  \centering
  \resizebox{\textwidth}{!}{\input{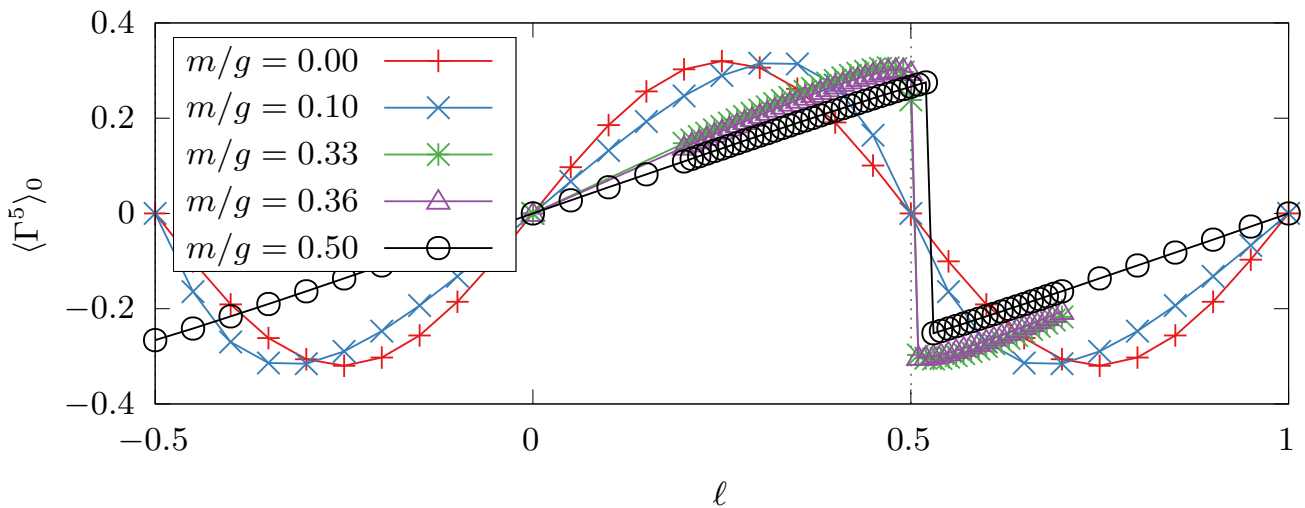}}
  \caption{$\langle \Gamma^5 \rangle_0$ versus $\ell$ for $x=1000$ and $m/g \in \{0,0.1,0.5\}$ in the range $\ell\in[-0.5,0.5]$, and for $m/g \in \{0.33,0.36\}$ in $\ell\in[0,0.7]$. Vertical dotted lines emphasise $\ell=0$ and $\ell=0.5$.}
  \label{gnu:gamma5Plot}
\end{figure*}

We continue to verify our model by calculating the ground state axial fermion density, an order parameter of the critical point first proposed by Creutz \cite{creutz_chiral_1995} and numerically calculated by \cite{byrnes_density_2002}. In the continuum, the axial fermion density in the ground state is defined as
\begin{align}
    \label{continuumGamma5}
    \langle \Gamma^5 \rangle_0 = \langle i \bar \psi \gamma^5 \psi \rangle_0,
\end{align}
where $\langle \hat O \rangle_0$ refers to the expectation value of some observable, $\hat O$ measured in the ground state. On a well-converged infinite two-site unit cell lattice, this order parameter can be written as\cite{byrnes_density_2002}
\begin{align}
    \label{g5Discrete}
    \langle \Gamma^5 \rangle_0 &= \frac{ig\sqrt{x}}{2} \left\langle \left(\phi^\dagger_0 \phi_{1} - \text{h.c}\right) - \left(\phi^\dagger_1 \phi_{2} - \text{h.c}\right) \right\rangle_0.
\end{align}

As we take $m\rightarrow 0$, the axial fermion density is expected to be proportionate to $\sin(\theta) \equiv \sin (2\pi \ell)$, as we see in Figure \ref{gnu:gamma5Plot}. However, above criticality, we expect a discontinuity at $\ell=0.5$, \cite{creutz_chiral_1995, byrnes_density_2002} and indeed that is what our numerical results reflect. The $m/g=0.33$ plot approaches $\left(\frac{m}{g}\right)_c$, but does not surpass criticality, leading to a sharp, but differentiable slope at $\ell=0.5$ which still passes through $\left\langle \Gamma^5 \right\rangle =0$. However, above the critical point, at $m/g=0.36$ and $m/g=0.5$, we see clear discontinuities, in line with expectations.

Finally, we explore evidence for the Schwinger model's characteristic `half-asymptotic' particles above the critical point. These half-asymptotic particles were first considered by Coleman, referring to particles which were deconfined, but only if ordered correctly \cite{coleman_more_1976}. To illustrate, consider the potential in the weak coupling limit, which Coleman approximates\footnote{Although an approximation, this term includes the full electrostatic potential and therefore the only origin of non-local forces \cite{byrnes_density_2002}.} to
\begin{align}
    V(r, \theta) = \frac{g^2}{2}\left(|r|-\frac{\theta}{\pi}r\right).
\end{align}
While particles in the model are generally confined, in the specific case of $|\theta| = \pi$, if $\theta$ is positive (negative) the potential drops to 0 for all positive (negative) $x$. Hence, in the case of $\theta = \pm k\pi$, particles can be deconfined, so long as they are correctly ordered. For a positive (negative) background field, any pair of fermionic and antifermionic particles is deconfined if the antifermion (fermion) is on the left.

As a result, we expect to see an asymmetry above the critical point when observing the correlation of a given antifermionic site occupation with the occupation of a fermionic site on its right versus its left. That is, we expect that for some fermionic site, $n$,
$\langle n^{(f)}_{n} n^{(f)}_{n+k} \rangle \neq \langle n^{(f)}_{n-k} n^{(f)}_{n} \rangle$ above the critical point.
 
\begin{figure*}[t]
  \centering
  \resizebox{\textwidth}{!}{\input{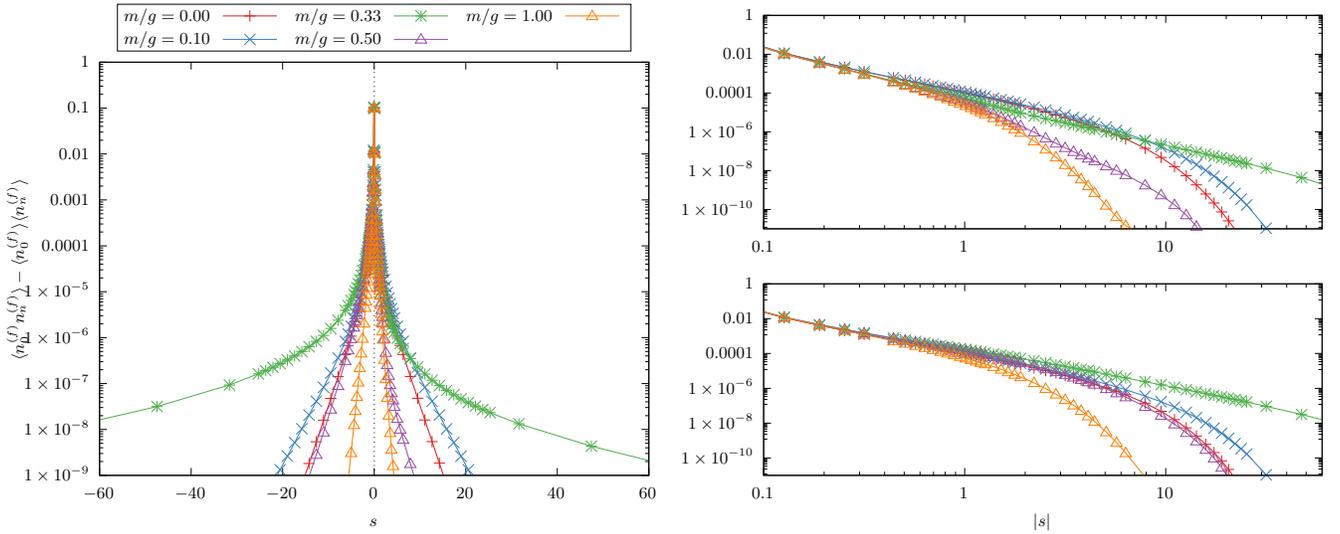}}
  \caption{The correlation of an antifermionic site occupation with fermionic site occupations plotted against unit length separation, $s=\frac{k}{\sqrt{x}}$, where $n$ is the fermion site number relative to the antifermion (left). When plotting the correlation for site separation to the right (top right) and left (bottom right) logarithmically, we identify a power-law relationship near the critical point $m/g\approx 0.33$, as expected.}
  \label{gnu:halfAsymptotic}
\end{figure*}

To verify this, we plot the correlator
\begin{align}
    \langle n^{(f)}_{n} n^{(f)}_{n+k} \rangle \neq \langle n^{(f)}_{n-k} n^{(f)}_{n} \rangle
\end{align}
for various masses and $\ell=0.5$ in Figure \ref{gnu:halfAsymptotic}. We find that for $m/g$ values below the critical point, we see symmetries around site separation $s=0$, while for values above criticality, we see an asymmetry (left). In particular, for $m/g=0.33$, which is close to the numerically calculated critical point, we see a power law relationship, further verified by switching to a logarithmic separation in \ref{gnu:halfAsymptotic} (right).

\section{Conclusion and outlook}
We have introduced a general iMPS/MPO construction for low-dimensional lattice gauge theories that (i) integrates out the gauge links in 1+1D, (ii) retains a tunable background field (topological $\theta$ term), and (iii) works directly in the thermodynamic limit with minimal changes to standard iDMRG. The key algorithmic ingredient is a symmetric recasting of the electric-field contribution and corresponding boundary-vector corrections for the effective Hamiltonian blocks, so that the long-range confining interaction is encoded by a compact, upper-triangular MPO while the iDMRG update itself remains unchanged. This delivers the efficiency of a pure-matter description without sacrificing the physics of background fields.
As a benchmark, we verified standard features of the Schwinger model, including the free-fermion limit, the expected features in the string tension and the symmetry breaking at $\ell\approx 0.5$;  the axial density $\langle \Gamma_5\rangle_0(\ell)$ follows the anticipated sinusoidal response at small mass with discontinuities above the critical point. This confirms that the boundary-term implementation reproduces the correct confining physics while operating directly at infinite size.

Beyond the specific benchmark, the construction is methodologically quite general. The confining potential arises from path counting in the FSM/MPO and can in fact be engineered to realize different separation dependences: in a genuine 1D gauge theory this potential is linear, but the MPO procedure itself would permit arbitrary polynomial. This suggests straightforward adaptations to other Abelian models with long-range constraints, and to quasi-two-dimensional geometries such as infinite cylinders, where one exploits translation invariance along the cylinder axis. The boundary prescription we employ is fully compatible with $U(1)$ symmetry sectors and with modern iDMRG implementations, making it easy to combine with bond-dimension control, symmetry enforcement, and related numerical improvements. In particular, the construction is consistent with infinite boundary conditions (IBC)\cite{phien_infinite_2012,phien_dynamical_2013,PhysRevB.88.155116,Zauner_2015}, which allow one to simulate finite regions embedded in an infinite environment and thereby study local excitations or defects without edge artifacts. It also integrates smoothly with the excitation ansatz (EA) for iMPS, which provides access to elementary excitations and dispersion relations on top of the infinite-system ground state. Taken together, these features make the approach not only efficient for ground-state studies, but also versatile for probing excitation spectra, quench dynamics, and response functions in lattice gauge theories.

\bibliography{Bibliography}

\begin{thebibliography}{40}%
\makeatletter
\providecommand \@ifxundefined [1]{%
 \@ifx{#1\undefined}
}%
\providecommand \@ifnum [1]{%
 \ifnum #1\expandafter \@firstoftwo
 \else \expandafter \@secondoftwo
 \fi
}%
\providecommand \@ifx [1]{%
 \ifx #1\expandafter \@firstoftwo
 \else \expandafter \@secondoftwo
 \fi
}%
\providecommand \natexlab [1]{#1}%
\providecommand \enquote  [1]{``#1''}%
\providecommand \bibnamefont  [1]{#1}%
\providecommand \bibfnamefont [1]{#1}%
\providecommand \citenamefont [1]{#1}%
\providecommand \href@noop [0]{\@secondoftwo}%
\providecommand \href [0]{\begingroup \@sanitize@url \@href}%
\providecommand \@href[1]{\@@startlink{#1}\@@href}%
\providecommand \@@href[1]{\endgroup#1\@@endlink}%
\providecommand \@sanitize@url [0]{\catcode `\\12\catcode `\$12\catcode
  `\&12\catcode `\#12\catcode `\^12\catcode `\_12\catcode `\%12\relax}%
\providecommand \@@startlink[1]{}%
\providecommand \@@endlink[0]{}%
\providecommand \url  [0]{\begingroup\@sanitize@url \@url }%
\providecommand \@url [1]{\endgroup\@href {#1}{\urlprefix }}%
\providecommand \urlprefix  [0]{URL }%
\providecommand \Eprint [0]{\href }%
\providecommand \doibase [0]{http://dx.doi.org/}%
\providecommand \selectlanguage [0]{\@gobble}%
\providecommand \bibinfo  [0]{\@secondoftwo}%
\providecommand \bibfield  [0]{\@secondoftwo}%
\providecommand \translation [1]{[#1]}%
\providecommand \BibitemOpen [0]{}%
\providecommand \bibitemStop [0]{}%
\providecommand \bibitemNoStop [0]{.\EOS\space}%
\providecommand \EOS [0]{\spacefactor3000\relax}%
\providecommand \BibitemShut  [1]{\csname bibitem#1\endcsname}%
\let\auto@bib@innerbib\@empty
\bibitem [{\citenamefont {Liu}\ \emph {et~al.}(2024)\citenamefont {Liu},
  \citenamefont {Zhang}, \citenamefont {Zhu}, \citenamefont {He}, \citenamefont
  {Yuan},\ and\ \citenamefont {Pan}}]{LiuStringBreaking}%
  \BibitemOpen
  \bibfield  {author} {\bibinfo {author} {\bibfnamefont {Ying}\ \bibnamefont
  {Liu}}, \bibinfo {author} {\bibfnamefont {Wei-Yong}\ \bibnamefont {Zhang}},
  \bibinfo {author} {\bibfnamefont {Zi-Hang}\ \bibnamefont {Zhu}}, \bibinfo
  {author} {\bibfnamefont {Ming-Gen}\ \bibnamefont {He}}, \bibinfo {author}
  {\bibfnamefont {Zhen-Sheng}\ \bibnamefont {Yuan}}, \ and\ \bibinfo {author}
  {\bibfnamefont {Jian-Wei}\ \bibnamefont {Pan}},\ }\bibfield  {title}
  {\enquote {\bibinfo {title} {String breaking mechanism in a lattice schwinger
  model simulator},}\ }\href {https://arxiv.org/abs/2411.15443} {\  (\bibinfo
  {year} {2024})},\ \Eprint {http://arxiv.org/abs/2411.15443} {arXiv:2411.15443
  [cond-mat.quant-gas]} \BibitemShut {NoStop}%
\bibitem [{\citenamefont {Kogut}\ and\ \citenamefont
  {Susskind}(1975)}]{kogut_hamiltonian_1975}%
  \BibitemOpen
  \bibfield  {author} {\bibinfo {author} {\bibfnamefont {John}\ \bibnamefont
  {Kogut}}\ and\ \bibinfo {author} {\bibfnamefont {Leonard}\ \bibnamefont
  {Susskind}},\ }\bibfield  {title} {\enquote {\bibinfo {title} {Hamiltonian
  formulation of {Wilson}'s lattice gauge theories},}\ }\href {\doibase
  10.1103/PhysRevD.11.395} {\bibfield  {journal} {\bibinfo  {journal} {Physical
  Review D}\ }\textbf {\bibinfo {volume} {11}},\ \bibinfo {pages} {395--408}
  (\bibinfo {year} {1975})},\ \bibinfo {note} {publisher: American Physical
  Society}\BibitemShut {NoStop}%
\bibitem [{\citenamefont {Byrnes}\ \emph {et~al.}(2002)\citenamefont {Byrnes},
  \citenamefont {Sriganesh}, \citenamefont {Bursill},\ and\ \citenamefont
  {Hamer}}]{byrnes_density_2002}%
  \BibitemOpen
  \bibfield  {author} {\bibinfo {author} {\bibfnamefont {T.~M.~R.}\
  \bibnamefont {Byrnes}}, \bibinfo {author} {\bibfnamefont {P.}~\bibnamefont
  {Sriganesh}}, \bibinfo {author} {\bibfnamefont {R.~J.}\ \bibnamefont
  {Bursill}}, \ and\ \bibinfo {author} {\bibfnamefont {C.~J.}\ \bibnamefont
  {Hamer}},\ }\bibfield  {title} {{\selectlanguage {en}\enquote {\bibinfo
  {title} {Density matrix renormalisation group approach to the massive
  {Schwinger} model},}\ }}\href {\doibase 10.1016/S0920-5632(02)01416-0}
  {\bibfield  {journal} {\bibinfo  {journal} {Nuclear Physics B - Proceedings
  Supplements}\ }\textbf {\bibinfo {volume} {109}},\ \bibinfo {pages}
  {202--206} (\bibinfo {year} {2002})}\BibitemShut {NoStop}%
\bibitem [{\citenamefont {Byrnes}(2003)}]{byrnes_density_2003}%
  \BibitemOpen
  \bibfield  {author} {\bibinfo {author} {\bibfnamefont {Tim}\ \bibnamefont
  {Byrnes}},\ }\emph {\bibinfo {title} {Density matrix renormalization group: a
  new approach to lattice gauge theory}},\ \href@noop {} {Ph.D. thesis},\
  \bibinfo  {school} {The University of New South Wales} (\bibinfo {year}
  {2003})\BibitemShut {NoStop}%
\bibitem [{\citenamefont {Bañuls}\ \emph {et~al.}(2013)\citenamefont
  {Bañuls}, \citenamefont {Cichy}, \citenamefont {Cirac},\ and\ \citenamefont
  {Jansen}}]{banuls_mass_2013}%
  \BibitemOpen
  \bibfield  {author} {\bibinfo {author} {\bibfnamefont {M.C.}\ \bibnamefont
  {Bañuls}}, \bibinfo {author} {\bibfnamefont {K.}~\bibnamefont {Cichy}},
  \bibinfo {author} {\bibfnamefont {J.I.}\ \bibnamefont {Cirac}}, \ and\
  \bibinfo {author} {\bibfnamefont {K.}~\bibnamefont {Jansen}},\ }\bibfield
  {title} {{\selectlanguage {en}\enquote {\bibinfo {title} {The mass spectrum
  of the {Schwinger} model with matrix product states},}\ }}\href {\doibase
  10.1007/JHEP11(2013)158} {\bibfield  {journal} {\bibinfo  {journal} {Journal
  of High Energy Physics}\ }\textbf {\bibinfo {volume} {2013}},\ \bibinfo
  {pages} {158} (\bibinfo {year} {2013})}\BibitemShut {NoStop}%
\bibitem [{\citenamefont {Bañuls}\ \emph {et~al.}(2017)\citenamefont
  {Bañuls}, \citenamefont {Cichy}, \citenamefont {Cirac}, \citenamefont
  {Jansen},\ and\ \citenamefont {Kühn}}]{banuls_density_2017}%
  \BibitemOpen
  \bibfield  {author} {\bibinfo {author} {\bibfnamefont {Mari~Carmen}\
  \bibnamefont {Bañuls}}, \bibinfo {author} {\bibfnamefont {Krzysztof}\
  \bibnamefont {Cichy}}, \bibinfo {author} {\bibfnamefont {J.~Ignacio}\
  \bibnamefont {Cirac}}, \bibinfo {author} {\bibfnamefont {Karl}\ \bibnamefont
  {Jansen}}, \ and\ \bibinfo {author} {\bibfnamefont {Stefan}\ \bibnamefont
  {Kühn}},\ }\bibfield  {title} {{\selectlanguage {eng}\enquote {\bibinfo
  {title} {Density {Induced} {Phase} {Transitions} in the {Schwinger} {Model}:
  {A} {Study} with {Matrix} {Product} {States}},}\ }}\href {\doibase
  10.1103/PhysRevLett.118.071601} {\bibfield  {journal} {\bibinfo  {journal}
  {Physical Review Letters}\ }\textbf {\bibinfo {volume} {118}},\ \bibinfo
  {pages} {071601} (\bibinfo {year} {2017})}\BibitemShut {NoStop}%
\bibitem [{\citenamefont {Buyens}\ \emph
  {et~al.}(2017{\natexlab{a}})\citenamefont {Buyens}, \citenamefont {Haegeman},
  \citenamefont {Hebenstreit}, \citenamefont {Verstraete},\ and\ \citenamefont
  {Van~Acoleyen}}]{buyens_real-time_2017}%
  \BibitemOpen
  \bibfield  {author} {\bibinfo {author} {\bibfnamefont {Boye}\ \bibnamefont
  {Buyens}}, \bibinfo {author} {\bibfnamefont {Jutho}\ \bibnamefont
  {Haegeman}}, \bibinfo {author} {\bibfnamefont {Florian}\ \bibnamefont
  {Hebenstreit}}, \bibinfo {author} {\bibfnamefont {Frank}\ \bibnamefont
  {Verstraete}}, \ and\ \bibinfo {author} {\bibfnamefont {Karel}\ \bibnamefont
  {Van~Acoleyen}},\ }\bibfield  {title} {\enquote {\bibinfo {title} {Real-time
  simulation of the {Schwinger} effect with matrix product states},}\ }\href
  {\doibase 10.1103/PhysRevD.96.114501} {\bibfield  {journal} {\bibinfo
  {journal} {Physical Review D}\ }\textbf {\bibinfo {volume} {96}},\ \bibinfo
  {pages} {114501} (\bibinfo {year} {2017}{\natexlab{a}})},\ \bibinfo {note}
  {publisher: American Physical Society}\BibitemShut {NoStop}%
\bibitem [{\citenamefont {Buyens}\ \emph
  {et~al.}(2017{\natexlab{b}})\citenamefont {Buyens}, \citenamefont
  {Montangero}, \citenamefont {Haegeman}, \citenamefont {Verstraete},\ and\
  \citenamefont {Van~Acoleyen}}]{buyens_finite-representation_2017}%
  \BibitemOpen
  \bibfield  {author} {\bibinfo {author} {\bibfnamefont {Boye}\ \bibnamefont
  {Buyens}}, \bibinfo {author} {\bibfnamefont {Simone}\ \bibnamefont
  {Montangero}}, \bibinfo {author} {\bibfnamefont {Jutho}\ \bibnamefont
  {Haegeman}}, \bibinfo {author} {\bibfnamefont {Frank}\ \bibnamefont
  {Verstraete}}, \ and\ \bibinfo {author} {\bibfnamefont {Karel}\ \bibnamefont
  {Van~Acoleyen}},\ }\bibfield  {title} {\enquote {\bibinfo {title}
  {Finite-representation approximation of lattice gauge theories at the
  continuum limit with tensor networks},}\ }\href {\doibase
  10.1103/PhysRevD.95.094509} {\bibfield  {journal} {\bibinfo  {journal}
  {Physical Review D}\ }\textbf {\bibinfo {volume} {95}},\ \bibinfo {pages}
  {094509} (\bibinfo {year} {2017}{\natexlab{b}})},\ \bibinfo {note}
  {publisher: American Physical Society}\BibitemShut {NoStop}%
\bibitem [{\citenamefont {Papaefstathiou}\ \emph {et~al.}(2021)\citenamefont
  {Papaefstathiou}, \citenamefont {Robaina}, \citenamefont {Cirac},\ and\
  \citenamefont {Bañuls}}]{papaefstathiou_density_2021}%
  \BibitemOpen
  \bibfield  {author} {\bibinfo {author} {\bibfnamefont {Irene}\ \bibnamefont
  {Papaefstathiou}}, \bibinfo {author} {\bibfnamefont {Daniel}\ \bibnamefont
  {Robaina}}, \bibinfo {author} {\bibfnamefont {J.~Ignacio}\ \bibnamefont
  {Cirac}}, \ and\ \bibinfo {author} {\bibfnamefont {Mari~Carmen}\ \bibnamefont
  {Bañuls}},\ }\bibfield  {title} {\enquote {\bibinfo {title} {Density of
  states of the lattice {Schwinger} model},}\ }\href {\doibase
  10.1103/PhysRevD.104.014514} {\bibfield  {journal} {\bibinfo  {journal}
  {Physical Review D}\ }\textbf {\bibinfo {volume} {104}},\ \bibinfo {pages}
  {014514} (\bibinfo {year} {2021})},\ \bibinfo {note} {publisher: American
  Physical Society}\BibitemShut {NoStop}%
\bibitem [{\citenamefont {Bañuls}\ and\ \citenamefont
  {Cichy}(2020)}]{banuls_review_2020}%
  \BibitemOpen
  \bibfield  {author} {\bibinfo {author} {\bibfnamefont {Mari~Carmen}\
  \bibnamefont {Bañuls}}\ and\ \bibinfo {author} {\bibfnamefont {Krzysztof}\
  \bibnamefont {Cichy}},\ }\bibfield  {title} {\enquote {\bibinfo {title}
  {Review on novel methods for lattice gauge theories},}\ }\href {\doibase
  10.1088/1361-6633/ab6311} {\bibfield  {journal} {\bibinfo  {journal} {Reports
  on Progress in Physics}\ }\textbf {\bibinfo {volume} {83}},\ \bibinfo {pages}
  {024401} (\bibinfo {year} {2020})},\ \bibinfo {note} {arXiv:
  1910.00257}\BibitemShut {NoStop}%
\bibitem [{\citenamefont {Magnifico}\ \emph {et~al.}(2025)\citenamefont
  {Magnifico}, \citenamefont {Cataldi}, \citenamefont {Rigobello},
  \citenamefont {Majcen}, \citenamefont {Jaschke}, \citenamefont {Silvi},\ and\
  \citenamefont {Montangero}}]{Magnifico2025TensorNetworks}%
  \BibitemOpen
  \bibfield  {author} {\bibinfo {author} {\bibfnamefont {Giuseppe}\
  \bibnamefont {Magnifico}}, \bibinfo {author} {\bibfnamefont {Giovanni}\
  \bibnamefont {Cataldi}}, \bibinfo {author} {\bibfnamefont {Marco}\
  \bibnamefont {Rigobello}}, \bibinfo {author} {\bibfnamefont {Peter}\
  \bibnamefont {Majcen}}, \bibinfo {author} {\bibfnamefont {Daniel}\
  \bibnamefont {Jaschke}}, \bibinfo {author} {\bibfnamefont {Pietro}\
  \bibnamefont {Silvi}}, \ and\ \bibinfo {author} {\bibfnamefont {Simone}\
  \bibnamefont {Montangero}},\ }\bibfield  {title} {\enquote {\bibinfo {title}
  {Tensor networks for lattice gauge theories beyond one dimension},}\ }\href
  {\doibase 10.1038/s42005-025-02125-x} {\bibfield  {journal} {\bibinfo
  {journal} {Communications Physics}\ }\textbf {\bibinfo {volume} {8}},\
  \bibinfo {pages} {322} (\bibinfo {year} {2025})}\BibitemShut {NoStop}%
\bibitem [{\citenamefont {Kelman}\ \emph {et~al.}(2024)\citenamefont {Kelman},
  \citenamefont {Borla}, \citenamefont {Gomelski}, \citenamefont {Elyovich},
  \citenamefont {Roose}, \citenamefont {Emonts},\ and\ \citenamefont
  {Zohar}}]{KelmanGaussian2024}%
  \BibitemOpen
  \bibfield  {author} {\bibinfo {author} {\bibfnamefont {Ariel}\ \bibnamefont
  {Kelman}}, \bibinfo {author} {\bibfnamefont {Umberto}\ \bibnamefont {Borla}},
  \bibinfo {author} {\bibfnamefont {Itay}\ \bibnamefont {Gomelski}}, \bibinfo
  {author} {\bibfnamefont {Jonathan}\ \bibnamefont {Elyovich}}, \bibinfo
  {author} {\bibfnamefont {Gertian}\ \bibnamefont {Roose}}, \bibinfo {author}
  {\bibfnamefont {Patrick}\ \bibnamefont {Emonts}}, \ and\ \bibinfo {author}
  {\bibfnamefont {Erez}\ \bibnamefont {Zohar}},\ }\bibfield  {title} {\enquote
  {\bibinfo {title} {Gauged gaussian projected entangled pair states: A high
  dimensional tensor network formulation for lattice gauge theories},}\ }\href
  {\doibase 10.1103/PhysRevD.110.054511} {\bibfield  {journal} {\bibinfo
  {journal} {Phys. Rev. D}\ }\textbf {\bibinfo {volume} {110}},\ \bibinfo
  {pages} {054511} (\bibinfo {year} {2024})}\BibitemShut {NoStop}%
\bibitem [{\citenamefont {Montangero}\ \emph {et~al.}(2021)\citenamefont
  {Montangero}, \citenamefont {Rico},\ and\ \citenamefont
  {Silvi}}]{Montangero2021LoopFreeTensor}%
  \BibitemOpen
  \bibfield  {author} {\bibinfo {author} {\bibfnamefont {S.}~\bibnamefont
  {Montangero}}, \bibinfo {author} {\bibfnamefont {E.}~\bibnamefont {Rico}}, \
  and\ \bibinfo {author} {\bibfnamefont {P.}~\bibnamefont {Silvi}},\ }\bibfield
   {title} {\enquote {\bibinfo {title} {Loop‐free tensor networks for
  high‐energy physics},}\ }\href {\doibase 10.1098/rsta.2021.0065} {\bibfield
   {journal} {\bibinfo  {journal} {Philosophical Transactions of the Royal
  Society A}\ }\textbf {\bibinfo {volume} {380}},\ \bibinfo {pages} {20210065}
  (\bibinfo {year} {2021})}\BibitemShut {NoStop}%
\bibitem [{\citenamefont {Zohar}(2022)}]{Zohar2022QuantumSimulationLGTs}%
  \BibitemOpen
  \bibfield  {author} {\bibinfo {author} {\bibfnamefont {Erez}\ \bibnamefont
  {Zohar}},\ }\bibfield  {title} {\enquote {\bibinfo {title} {Quantum
  simulation of lattice gauge theories in more than one space
  dimension-–requirements, challenges and methods},}\ }\href {\doibase
  10.1098/rsta.2021.0069} {\bibfield  {journal} {\bibinfo  {journal}
  {Philosophical Transactions of the Royal Society A: Mathematical, Physical
  and Engineering Sciences}\ }\textbf {\bibinfo {volume} {380}},\ \bibinfo
  {pages} {20210069} (\bibinfo {year} {2022})},\ \bibinfo {note} {epub 20
  December 2021}\BibitemShut {NoStop}%
\bibitem [{\citenamefont {Aidelsburger}\ \emph {et~al.}(2022)\citenamefont
  {Aidelsburger}, \citenamefont {Barbiero}, \citenamefont {Bermudez},
  \citenamefont {Chanda}, \citenamefont {Dauphin}, \citenamefont
  {Gonz{\'a}lez-Cuadra}, \citenamefont {Grzybowski}, \citenamefont {Hands},
  \citenamefont {Jendrzejewski}, \citenamefont {J{\"u}nemann}, \citenamefont
  {Juzeli{\=u}nas}, \citenamefont {Kasper}, \citenamefont {Piga}, \citenamefont
  {Ran}, \citenamefont {Rizzi}, \citenamefont {Sierra}, \citenamefont
  {Tagliacozzo}, \citenamefont {Tirrito}, \citenamefont {Zache}, \citenamefont
  {Zakrzewski}, \citenamefont {Zohar},\ and\ \citenamefont
  {Lewenstein}}]{Aidelsburger2022ColdAtomsLGT}%
  \BibitemOpen
  \bibfield  {author} {\bibinfo {author} {\bibfnamefont {Monika}\ \bibnamefont
  {Aidelsburger}}, \bibinfo {author} {\bibfnamefont {Luca}\ \bibnamefont
  {Barbiero}}, \bibinfo {author} {\bibfnamefont {Alejandro}\ \bibnamefont
  {Bermudez}}, \bibinfo {author} {\bibfnamefont {Titas}\ \bibnamefont
  {Chanda}}, \bibinfo {author} {\bibfnamefont {Alexandre}\ \bibnamefont
  {Dauphin}}, \bibinfo {author} {\bibfnamefont {Daniel}\ \bibnamefont
  {Gonz{\'a}lez-Cuadra}}, \bibinfo {author} {\bibfnamefont {Piotr~R.}\
  \bibnamefont {Grzybowski}}, \bibinfo {author} {\bibfnamefont {Simon}\
  \bibnamefont {Hands}}, \bibinfo {author} {\bibfnamefont {Fred}\ \bibnamefont
  {Jendrzejewski}}, \bibinfo {author} {\bibfnamefont {Johannes}\ \bibnamefont
  {J{\"u}nemann}}, \bibinfo {author} {\bibfnamefont {Gediminas}\ \bibnamefont
  {Juzeli{\=u}nas}}, \bibinfo {author} {\bibfnamefont {Valentin}\ \bibnamefont
  {Kasper}}, \bibinfo {author} {\bibfnamefont {Angel}\ \bibnamefont {Piga}},
  \bibinfo {author} {\bibfnamefont {Shi-Ju}\ \bibnamefont {Ran}}, \bibinfo
  {author} {\bibfnamefont {Matteo}\ \bibnamefont {Rizzi}}, \bibinfo {author}
  {\bibfnamefont {Germ{\'a}n}\ \bibnamefont {Sierra}}, \bibinfo {author}
  {\bibfnamefont {Luca}\ \bibnamefont {Tagliacozzo}}, \bibinfo {author}
  {\bibfnamefont {Emanuele}\ \bibnamefont {Tirrito}}, \bibinfo {author}
  {\bibfnamefont {Torsten~V.}\ \bibnamefont {Zache}}, \bibinfo {author}
  {\bibfnamefont {Jakub}\ \bibnamefont {Zakrzewski}}, \bibinfo {author}
  {\bibfnamefont {Erez}\ \bibnamefont {Zohar}}, \ and\ \bibinfo {author}
  {\bibfnamefont {Maciej}\ \bibnamefont {Lewenstein}},\ }\bibfield  {title}
  {\enquote {\bibinfo {title} {Cold atoms meet lattice gauge theory},}\ }\href
  {\doibase 10.1098/rsta.2021.0064} {\bibfield  {journal} {\bibinfo  {journal}
  {Philosophical Transactions of the Royal Society A: Mathematical, Physical
  and Engineering Sciences}\ }\textbf {\bibinfo {volume} {380}},\ \bibinfo
  {pages} {20210064} (\bibinfo {year} {2022})},\ \bibinfo {note} {epub 20
  December 2021}\BibitemShut {NoStop}%
\bibitem [{\citenamefont {Gleis}\ \emph {et~al.}(2023)\citenamefont {Gleis},
  \citenamefont {Li},\ and\ \citenamefont {von
  Delft}}]{PhysRevLett.130.246402}%
  \BibitemOpen
  \bibfield  {author} {\bibinfo {author} {\bibfnamefont {Andreas}\ \bibnamefont
  {Gleis}}, \bibinfo {author} {\bibfnamefont {Jheng-Wei}\ \bibnamefont {Li}}, \
  and\ \bibinfo {author} {\bibfnamefont {Jan}\ \bibnamefont {von Delft}},\
  }\bibfield  {title} {\enquote {\bibinfo {title} {Controlled bond expansion
  for density matrix renormalization group ground state search at single-site
  costs},}\ }\href {\doibase 10.1103/PhysRevLett.130.246402} {\bibfield
  {journal} {\bibinfo  {journal} {Phys. Rev. Lett.}\ }\textbf {\bibinfo
  {volume} {130}},\ \bibinfo {pages} {246402} (\bibinfo {year}
  {2023})}\BibitemShut {NoStop}%
\bibitem [{\citenamefont {McCulloch}\ and\ \citenamefont
  {Osborne}(2024)}]{CBEComment}%
  \BibitemOpen
  \bibfield  {author} {\bibinfo {author} {\bibfnamefont {Ian~P}\ \bibnamefont
  {McCulloch}}\ and\ \bibinfo {author} {\bibfnamefont {Jesse~J}\ \bibnamefont
  {Osborne}},\ }\href {https://arxiv.org/abs/2403.00562} {\enquote {\bibinfo
  {title} {Comment on "controlled bond expansion for density matrix
  renormalization group ground state search at single-site costs" (extended
  version)},}\ } (\bibinfo {year} {2024}),\ \Eprint
  {http://arxiv.org/abs/2403.00562} {arXiv:2403.00562 [cond-mat.str-el]}
  \BibitemShut {NoStop}%
\bibitem [{\citenamefont {Banks}\ \emph {et~al.}(1976)\citenamefont {Banks},
  \citenamefont {Susskind},\ and\ \citenamefont {Kogut}}]{staggeredFermions}%
  \BibitemOpen
  \bibfield  {author} {\bibinfo {author} {\bibfnamefont {T.}~\bibnamefont
  {Banks}}, \bibinfo {author} {\bibfnamefont {Leonard}\ \bibnamefont
  {Susskind}}, \ and\ \bibinfo {author} {\bibfnamefont {John}\ \bibnamefont
  {Kogut}},\ }\bibfield  {title} {\enquote {\bibinfo {title} {Strong-coupling
  calculations of lattice gauge theories: (1 + 1)-dimensional exercises},}\
  }\href {\doibase 10.1103/PhysRevD.13.1043} {\bibfield  {journal} {\bibinfo
  {journal} {Physical Review D}\ }\textbf {\bibinfo {volume} {13}},\ \bibinfo
  {pages} {1043--1053} (\bibinfo {year} {1976})},\ \bibinfo {note} {publisher:
  American Physical Society}\BibitemShut {NoStop}%
\bibitem [{\citenamefont {Szyniszewski}(2013)}]{szyniszewski_numerical_2013}%
  \BibitemOpen
  \bibfield  {author} {\bibinfo {author} {\bibfnamefont {Marcin}\ \bibnamefont
  {Szyniszewski}},\ }\emph {\bibinfo {title} {Numerical investigations of the
  {Schwinger} model and selected quantum spin models}},\ \href@noop {}
  {Master's thesis},\ \bibinfo  {school} {Adam Mickiewicz University}, \bibinfo
  {address} {Poznań, Poland} (\bibinfo {year} {2013})\BibitemShut {NoStop}%
\bibitem [{\citenamefont {Susskind}(1977)}]{susskind_lattice_1977}%
  \BibitemOpen
  \bibfield  {author} {\bibinfo {author} {\bibfnamefont {Leonard}\ \bibnamefont
  {Susskind}},\ }\bibfield  {title} {\enquote {\bibinfo {title} {Lattice
  fermions},}\ }\href {\doibase 10.1103/PhysRevD.16.3031} {\bibfield  {journal}
  {\bibinfo  {journal} {Physical Review D}\ }\textbf {\bibinfo {volume} {16}},\
  \bibinfo {pages} {3031--3039} (\bibinfo {year} {1977})},\ \bibinfo {note}
  {publisher: American Physical Society}\BibitemShut {NoStop}%
\bibitem [{\citenamefont {Bodwin}\ and\ \citenamefont
  {Kovacs}(1987)}]{bodwin_lattice_1987}%
  \BibitemOpen
  \bibfield  {author} {\bibinfo {author} {\bibfnamefont {Geoffrey~T.}\
  \bibnamefont {Bodwin}}\ and\ \bibinfo {author} {\bibfnamefont {Eve~V.}\
  \bibnamefont {Kovacs}},\ }\bibfield  {title} {\enquote {\bibinfo {title}
  {Lattice fermions in the {Schwinger} model},}\ }\href {\doibase
  10.1103/PhysRevD.35.3198} {\bibfield  {journal} {\bibinfo  {journal}
  {Physical Review D}\ }\textbf {\bibinfo {volume} {35}},\ \bibinfo {pages}
  {3198--3218} (\bibinfo {year} {1987})},\ \bibinfo {note} {publisher: American
  Physical Society}\BibitemShut {NoStop}%
\bibitem [{\citenamefont {Zohar}\ and\ \citenamefont
  {Burrello}(2015)}]{zohar_formulation_2015}%
  \BibitemOpen
  \bibfield  {author} {\bibinfo {author} {\bibfnamefont {Erez}\ \bibnamefont
  {Zohar}}\ and\ \bibinfo {author} {\bibfnamefont {Michele}\ \bibnamefont
  {Burrello}},\ }\bibfield  {title} {\enquote {\bibinfo {title} {Formulation of
  lattice gauge theories for quantum simulations},}\ }\href {\doibase
  10.1103/PhysRevD.91.054506} {\bibfield  {journal} {\bibinfo  {journal}
  {Physical Review D}\ }\textbf {\bibinfo {volume} {91}},\ \bibinfo {pages}
  {054506} (\bibinfo {year} {2015})},\ \bibinfo {note} {publisher: American
  Physical Society}\BibitemShut {NoStop}%
\bibitem [{\citenamefont {Kogut}(1978)}]{kogut_three_1978}%
  \BibitemOpen
  \bibfield  {author} {\bibinfo {author} {\bibfnamefont {J.~B.}\ \bibnamefont
  {Kogut}},\ }\bibfield  {title} {{\selectlanguage {en}\enquote {\bibinfo
  {title} {Three {Lectures} on {Lattice} {Gauge} {Theory}},}\ }}in\ \href
  {\doibase 10.1007/978-1-4684-2814-8_9} {{\selectlanguage {en}\emph {\bibinfo
  {booktitle} {Many {Degrees} of {Freedom} in {Particle} {Theory}}}}},\
  \bibinfo {series and number} {{NATO} {Advanced} {Study} {Institutes}
  {Series}},\ \bibinfo {editor} {edited by\ \bibinfo {editor} {\bibfnamefont
  {H.}~\bibnamefont {Satz}}}\ (\bibinfo  {publisher} {Springer US},\ \bibinfo
  {address} {Boston, MA},\ \bibinfo {year} {1978})\ pp.\ \bibinfo {pages}
  {275--343}\BibitemShut {NoStop}%
\bibitem [{\citenamefont {Hamer}\ \emph {et~al.}(1997)\citenamefont {Hamer},
  \citenamefont {Weihong},\ and\ \citenamefont {Oitmaa}}]{hamer_series_1997}%
  \BibitemOpen
  \bibfield  {author} {\bibinfo {author} {\bibfnamefont {C.~J.}\ \bibnamefont
  {Hamer}}, \bibinfo {author} {\bibfnamefont {Zheng}\ \bibnamefont {Weihong}},
  \ and\ \bibinfo {author} {\bibfnamefont {J.}~\bibnamefont {Oitmaa}},\
  }\bibfield  {title} {\enquote {\bibinfo {title} {Series {Expansions} for the
  {Massive} {Schwinger} {Model} in {Hamiltonian} lattice theory},}\ }\href
  {\doibase 10.1103/PhysRevD.56.55} {\bibfield  {journal} {\bibinfo  {journal}
  {Physical Review D}\ }\textbf {\bibinfo {volume} {56}},\ \bibinfo {pages}
  {55--67} (\bibinfo {year} {1997})},\ \bibinfo {note}
  {arXiv:hep-lat/9701015}\BibitemShut {NoStop}%
\bibitem [{\citenamefont {Zapp}\ and\ \citenamefont
  {Orús}(2017)}]{zapp_tensor_2017}%
  \BibitemOpen
  \bibfield  {author} {\bibinfo {author} {\bibfnamefont {Kai}\ \bibnamefont
  {Zapp}}\ and\ \bibinfo {author} {\bibfnamefont {Román}\ \bibnamefont
  {Orús}},\ }\bibfield  {title} {\enquote {\bibinfo {title} {Tensor network
  simulation of {QED} on infinite lattices: {Learning} from (1 +1 ) d , and
  prospects for (2 +1 ) d},}\ }\href {\doibase 10.1103/PhysRevD.95.114508}
  {\bibfield  {journal} {\bibinfo  {journal} {Physical Review D}\ }\textbf
  {\bibinfo {volume} {95}},\ \bibinfo {pages} {114508} (\bibinfo {year}
  {2017})},\ \bibinfo {note} {aDS Bibcode: 2017PhRvD..95k4508Z}\BibitemShut
  {NoStop}%
\bibitem [{\citenamefont {Schollwöck}(2005)}]{schollwock_density-matrix_2005}%
  \BibitemOpen
  \bibfield  {author} {\bibinfo {author} {\bibfnamefont {U.}~\bibnamefont
  {Schollwöck}},\ }\bibfield  {title} {\enquote {\bibinfo {title} {The
  density-matrix renormalization group},}\ }\href {\doibase
  10.1103/RevModPhys.77.259} {\bibfield  {journal} {\bibinfo  {journal}
  {Reviews of Modern Physics}\ }\textbf {\bibinfo {volume} {77}},\ \bibinfo
  {pages} {259--315} (\bibinfo {year} {2005})},\ \bibinfo {note} {publisher:
  American Physical Society}\BibitemShut {NoStop}%
\bibitem [{\citenamefont {McCulloch}(2008)}]{mcculloch_infinite_2008}%
  \BibitemOpen
  \bibfield  {author} {\bibinfo {author} {\bibfnamefont {I.~P.}\ \bibnamefont
  {McCulloch}},\ }\href {http://arxiv.org/abs/0804.2509} {\enquote {\bibinfo
  {title} {Infinite size density matrix renormalization group, revisited},}\ }
  (\bibinfo {year} {2008}),\ \bibinfo {note} {arXiv:0804.2509
  [cond-mat]}\BibitemShut {NoStop}%
\bibitem [{\citenamefont {Phien}\ \emph {et~al.}(2012)\citenamefont {Phien},
  \citenamefont {Vidal},\ and\ \citenamefont
  {McCulloch}}]{phien_infinite_2012}%
  \BibitemOpen
  \bibfield  {author} {\bibinfo {author} {\bibfnamefont {Ho~N.}\ \bibnamefont
  {Phien}}, \bibinfo {author} {\bibfnamefont {Guifré}\ \bibnamefont {Vidal}},
  \ and\ \bibinfo {author} {\bibfnamefont {Ian~P.}\ \bibnamefont {McCulloch}},\
  }\bibfield  {title} {\enquote {\bibinfo {title} {Infinite boundary conditions
  for matrix product state calculations},}\ }\href {\doibase
  10.1103/PhysRevB.86.245107} {\bibfield  {journal} {\bibinfo  {journal}
  {Physical Review B}\ }\textbf {\bibinfo {volume} {86}},\ \bibinfo {pages}
  {245107} (\bibinfo {year} {2012})},\ \bibinfo {note} {publisher: American
  Physical Society}\BibitemShut {NoStop}%
\bibitem [{\citenamefont {Michel}\ and\ \citenamefont
  {McCulloch}(2010)}]{michel_schur_2010}%
  \BibitemOpen
  \bibfield  {author} {\bibinfo {author} {\bibfnamefont {L.}~\bibnamefont
  {Michel}}\ and\ \bibinfo {author} {\bibfnamefont {I.~P.}\ \bibnamefont
  {McCulloch}},\ }\href {\doibase 10.48550/arXiv.1008.4667} {\enquote {\bibinfo
  {title} {Schur {Forms} of {Matrix} {Product} {Operators} in the {Infinite}
  {Limit}},}\ } (\bibinfo {year} {2010}),\ \bibinfo {note} {arXiv:1008.4667
  [cond-mat]}\BibitemShut {NoStop}%
\bibitem [{\citenamefont {Crewther}\ and\ \citenamefont
  {Hamer}(1980)}]{crewther_eigenvalues_1980}%
  \BibitemOpen
  \bibfield  {author} {\bibinfo {author} {\bibfnamefont {D.~P.}\ \bibnamefont
  {Crewther}}\ and\ \bibinfo {author} {\bibfnamefont {C.~J.}\ \bibnamefont
  {Hamer}},\ }\bibfield  {title} {{\selectlanguage {en}\enquote {\bibinfo
  {title} {Eigenvalues for the massive {Schwinger} model from a finite-lattice
  {Hamiltonian} approach},}\ }}\href {\doibase 10.1016/0550-3213(80)90154-6}
  {\bibfield  {journal} {\bibinfo  {journal} {Nuclear Physics B}\ }\bibinfo
  {series} {Volume {B170} [{FSI}] {No}. 3 to follow in {Approximately} {Two}
  {Months}},\ \textbf {\bibinfo {volume} {170}},\ \bibinfo {pages} {353--368}
  (\bibinfo {year} {1980})}\BibitemShut {NoStop}%
\bibitem [{\citenamefont {Mattis}(1981)}]{mattis_history_1981}%
  \BibitemOpen
  \bibfield  {author} {\bibinfo {author} {\bibfnamefont {Daniel~C.}\
  \bibnamefont {Mattis}},\ }\bibfield  {title} {{\selectlanguage {en}\enquote
  {\bibinfo {title} {History of {Magnetism}},}\ }}in\ \href {\doibase
  10.1007/978-3-642-83238-3_1} {{\selectlanguage {en}\emph {\bibinfo
  {booktitle} {The {Theory} of {Magnetism} {I}: {Statics} and {Dynamics}}}}},\
  \bibinfo {series and number} {Springer {Series} in {Solid}-{State}
  {Sciences}},\ \bibinfo {editor} {edited by\ \bibinfo {editor} {\bibfnamefont
  {Daniel~C.}\ \bibnamefont {Mattis}}}\ (\bibinfo  {publisher} {Springer},\
  \bibinfo {address} {Berlin, Heidelberg},\ \bibinfo {year} {1981})\ pp.\
  \bibinfo {pages} {1--38}\BibitemShut {NoStop}%
\bibitem [{\citenamefont {Fujii}\ \emph {et~al.}(2025)\citenamefont {Fujii},
  \citenamefont {Fujikura}, \citenamefont {Kikukawa}, \citenamefont {Okuda},\
  and\ \citenamefont {Pedersen}}]{Fujii2025Critical}%
  \BibitemOpen
  \bibfield  {author} {\bibinfo {author} {\bibfnamefont {Hirotsugu}\
  \bibnamefont {Fujii}}, \bibinfo {author} {\bibfnamefont {Kohei}\ \bibnamefont
  {Fujikura}}, \bibinfo {author} {\bibfnamefont {Yoshio}\ \bibnamefont
  {Kikukawa}}, \bibinfo {author} {\bibfnamefont {Takuya}\ \bibnamefont
  {Okuda}}, \ and\ \bibinfo {author} {\bibfnamefont {Juan~W.}\ \bibnamefont
  {Pedersen}},\ }\bibfield  {title} {\enquote {\bibinfo {title} {Critical
  behavior of the schwinger model via gauge-invariant variational uniform
  matrix product states},}\ }\href {\doibase 10.1103/PhysRevD.111.094505}
  {\bibfield  {journal} {\bibinfo  {journal} {Physical Review D}\ }\textbf
  {\bibinfo {volume} {111}},\ \bibinfo {pages} {094505} (\bibinfo {year}
  {2025})},\ \bibinfo {note} {received 21 February 2025; accepted 14 April
  2025; published 6 May 2025}\BibitemShut {NoStop}%
\bibitem [{\citenamefont {Hamer}\ \emph {et~al.}(1982)\citenamefont {Hamer},
  \citenamefont {Kogut}, \citenamefont {Crewther},\ and\ \citenamefont
  {Mazzolini}}]{hamer_massive_1982}%
  \BibitemOpen
  \bibfield  {author} {\bibinfo {author} {\bibfnamefont {C.~J.}\ \bibnamefont
  {Hamer}}, \bibinfo {author} {\bibfnamefont {J.}~\bibnamefont {Kogut}},
  \bibinfo {author} {\bibfnamefont {D.~P.}\ \bibnamefont {Crewther}}, \ and\
  \bibinfo {author} {\bibfnamefont {M.~M.}\ \bibnamefont {Mazzolini}},\
  }\bibfield  {title} {{\selectlanguage {en}\enquote {\bibinfo {title} {The
  massive {Schwinger} model on a lattice: {Background} field, chiral symmetry
  and the string tension},}\ }}\href {\doibase 10.1016/0550-3213(82)90229-2}
  {\bibfield  {journal} {\bibinfo  {journal} {Nuclear Physics B}\ }\textbf
  {\bibinfo {volume} {208}},\ \bibinfo {pages} {413--438} (\bibinfo {year}
  {1982})}\BibitemShut {NoStop}%
\bibitem [{\citenamefont {Coleman}(1976)}]{coleman_more_1976}%
  \BibitemOpen
  \bibfield  {author} {\bibinfo {author} {\bibfnamefont {Sidney}\ \bibnamefont
  {Coleman}},\ }\bibfield  {title} {{\selectlanguage {en}\enquote {\bibinfo
  {title} {More about the massive {Schwinger} model},}\ }}\href {\doibase
  10.1016/0003-4916(76)90280-3} {\bibfield  {journal} {\bibinfo  {journal}
  {Annals of Physics}\ }\textbf {\bibinfo {volume} {101}},\ \bibinfo {pages}
  {239--267} (\bibinfo {year} {1976})}\BibitemShut {NoStop}%
\bibitem [{\citenamefont {Schollwöck}(2011)}]{schollwock_density-matrix_2011}%
  \BibitemOpen
  \bibfield  {author} {\bibinfo {author} {\bibfnamefont {Ulrich}\ \bibnamefont
  {Schollwöck}},\ }\bibfield  {title} {{\selectlanguage {en}\enquote {\bibinfo
  {title} {The density-matrix renormalization group in the age of matrix
  product states},}\ }}\href {\doibase 10.1016/j.aop.2010.09.012} {\bibfield
  {journal} {\bibinfo  {journal} {Annals of Physics}\ }\bibinfo {series}
  {January 2011 {Special} {Issue}},\ \textbf {\bibinfo {volume} {326}},\
  \bibinfo {pages} {96--192} (\bibinfo {year} {2011})}\BibitemShut {NoStop}%
\bibitem [{\citenamefont {Creutz}(1995)}]{creutz_chiral_1995}%
  \BibitemOpen
  \bibfield  {author} {\bibinfo {author} {\bibfnamefont {Michael}\ \bibnamefont
  {Creutz}},\ }\bibfield  {title} {{\selectlanguage {en}\enquote {\bibinfo
  {title} {Chiral symmetry on the lattice},}\ }}\href {\doibase
  10.1016/0920-5632(95)00187-E} {\bibfield  {journal} {\bibinfo  {journal}
  {Nuclear Physics B - Proceedings Supplements}\ }\textbf {\bibinfo {volume}
  {42}},\ \bibinfo {pages} {56--66} (\bibinfo {year} {1995})}\BibitemShut
  {NoStop}%
\bibitem [{Note1()}]{Note1}%
  \BibitemOpen
  \bibinfo {note} {Although an approximation, this term includes the full
  electrostatic potential and therefore the only origin of non-local forces
  \cite {byrnes_density_2002}.}\BibitemShut {Stop}%
\bibitem [{\citenamefont {Phien}\ \emph {et~al.}(2013)\citenamefont {Phien},
  \citenamefont {Vidal},\ and\ \citenamefont
  {McCulloch}}]{phien_dynamical_2013}%
  \BibitemOpen
  \bibfield  {author} {\bibinfo {author} {\bibfnamefont {Ho~N.}\ \bibnamefont
  {Phien}}, \bibinfo {author} {\bibfnamefont {Guifr\'e}\ \bibnamefont {Vidal}},
  \ and\ \bibinfo {author} {\bibfnamefont {Ian~P.}\ \bibnamefont {McCulloch}},\
  }\bibfield  {title} {\enquote {\bibinfo {title} {Dynamical windows for
  real-time evolution with matrix product states},}\ }\href {\doibase
  10.1103/PhysRevB.88.035103} {\bibfield  {journal} {\bibinfo  {journal} {Phys.
  Rev. B}\ }\textbf {\bibinfo {volume} {88}},\ \bibinfo {pages} {035103}
  (\bibinfo {year} {2013})}\BibitemShut {NoStop}%
\bibitem [{\citenamefont {Milsted}\ \emph {et~al.}(2013)\citenamefont
  {Milsted}, \citenamefont {Haegeman}, \citenamefont {Osborne},\ and\
  \citenamefont {Verstraete}}]{PhysRevB.88.155116}%
  \BibitemOpen
  \bibfield  {author} {\bibinfo {author} {\bibfnamefont {Ashley}\ \bibnamefont
  {Milsted}}, \bibinfo {author} {\bibfnamefont {Jutho}\ \bibnamefont
  {Haegeman}}, \bibinfo {author} {\bibfnamefont {Tobias~J.}\ \bibnamefont
  {Osborne}}, \ and\ \bibinfo {author} {\bibfnamefont {Frank}\ \bibnamefont
  {Verstraete}},\ }\bibfield  {title} {\enquote {\bibinfo {title} {Variational
  matrix product ansatz for nonuniform dynamics in the thermodynamic limit},}\
  }\href {\doibase 10.1103/PhysRevB.88.155116} {\bibfield  {journal} {\bibinfo
  {journal} {Phys. Rev. B}\ }\textbf {\bibinfo {volume} {88}},\ \bibinfo
  {pages} {155116} (\bibinfo {year} {2013})}\BibitemShut {NoStop}%
\bibitem [{\citenamefont {Zauner}\ \emph {et~al.}(2015)\citenamefont {Zauner},
  \citenamefont {Ganahl}, \citenamefont {Evertz},\ and\ \citenamefont
  {Nishino}}]{Zauner_2015}%
  \BibitemOpen
  \bibfield  {author} {\bibinfo {author} {\bibfnamefont {V}~\bibnamefont
  {Zauner}}, \bibinfo {author} {\bibfnamefont {M}~\bibnamefont {Ganahl}},
  \bibinfo {author} {\bibfnamefont {H~G}\ \bibnamefont {Evertz}}, \ and\
  \bibinfo {author} {\bibfnamefont {T}~\bibnamefont {Nishino}},\ }\bibfield
  {title} {\enquote {\bibinfo {title} {Time evolution within a comoving window:
  scaling of signal fronts and magnetization plateaus after a local quench in
  quantum spin chains},}\ }\href {\doibase 10.1088/0953-8984/27/42/425602}
  {\bibfield  {journal} {\bibinfo  {journal} {Journal of Physics: Condensed
  Matter}\ }\textbf {\bibinfo {volume} {27}},\ \bibinfo {pages} {425602}
  (\bibinfo {year} {2015})}\BibitemShut {NoStop}%
\end{thebibliography}%

\appendix
\section{Proof of Generalised MPO}
\label{appendix:FSMProof}
Equation \eqref{polynomialMPO_Nn}'s validity can be confirmed via an inductive proof. Beginning with the base case, we simply seek to show that $\bigcirc_{n=0}^{\bar{N}} W^{[n]}$ reduces to Equation \eqref{polynomialMPO}'s $W$ for $\bar{N}=0$. We trivially see that this is the case along the diagonal, as, for both matrices, $W^{[0]} = W_{a,a} = \mathbb{I}, \; \forall a$. Next, we verify row-by-row using base-0 indexing.

In row $0$, we find that columns $1$ through to $p$ are characterised as
\begin{align}
    \bigcirc_{n=0}^{\bar{N}} W^{[n]}_{0,a} = \pihat{0}{a-1} \sum_{k=0}^{\bar{N}} \binom{\bar{N}-k}{a-1} \hat{Q}_k.
\end{align}
Setting $\bar{N}=0$, the upper index of the binomial coefficient in this expression is always $0$. For column $1$, the binomial coefficient equates to $\binom{0}{0} = 1$, and we are left with
\begin{align}
    W^{[0]}_{0,1} = \pihat{0}{0} \hat{Q}_0 = \hat{Q}_0,
\end{align}
as required. For columns $1$ to $p$, the upper index of the binomial coefficient is always less than the lower index, and hence the entire expression goes to $0$, as required. Column $p+1$ similarly reduces to $\hat{Q}_0$ as required, as $\binom{\bar{N}-k}{m}$ goes to $0$ where $\bar{N}=0$ and $0 \leq k \leq \bar{N}$. By a similar argument, column $p+2$ goes to $0$.

The upper-triangular cells of rows $1$ through $p$ can be characterised as $W^{[\bar{N}]}_{a,b} = \binom{N}{b-a} \pihat{a-1}{b-1}$ for columns up to and incuding $p$. For $\bar{N}=0$, this will only result in non-zero entries for $0 \leq b-a \leq 1$, as required. In the upper-triangular component of the matrix, this is only the case where $b = a+1$, which gives us
\begin{align}
    W^{[0]}_{a,a+1} = \binom{1}{1} \pihat{a-1}{a} = \sigma_{a-1}^{a},
\end{align}
as required, or where $b = a$, which gives us the diagonal identity entries as required. In the case of column $p+1$, we find that the lower index of the binomial coefficient, $k+1$, will always be greater than the upper coefficient when $k > \bar{N}$. Hence, for the base case of $\bar{N}=0$ we only consider the $k=0$ terms, and are left with
\begin{align}
    W^{[0]}_{a,p+1} = \binom{1}{1} \pihat{a-1}{a-1} \sigma_{a-1}^{p} = \sigma_{a-1}^{p},
\end{align}
as required. For column $p+2$ of row $1$, the inner sum, $\sum_{m=1}^k$, goes to zero when $\bar{N}=0$, as $k$ is upper bound by $\bar{N}$, leaving us only with $\hat{Q}_0$. For similar reasons, column $p+2$ in rows $2$ to $p$ all go to zero when $\bar{N}=0$. Finally, it is trivial to see that $W^{[0]}_{p+1,p+2} = \hat{Q}_0$ when $\bar{N}=0$, completing our base-case proof.

We now complete the inductive step, assuming our claim holds for $\bar{N}-1$. As a consequence of the upper-triangular structure of our MPO, we trivially see that our claim holds for the diagonal components:
\begin{align}
    \left(\bigcirc_{n=0}^{\bar{N}-1} W^{[n]}\right)_{a,a} W^{[\bar{N}]}_{a,a} = \mathbb{I}.
\end{align}
For the top row, we find that cells $ \left(\bigcirc_{n=0}^{\bar{N}} W^{[n]}\right)_{0,a}$ for $2 \leq a \leq p$ are equal to
\begin{widetext}
    \begin{align}
    \left(\bigcirc_{n=0}^{\bar{N}} W^{[n]}\right)_{0,a} &= \left(\bigcirc_{n=0}^{\bar{N}-1} W^{[n]}\right)_{0,a} W^{[\bar{N}]}_{a,a} + \left(\bigcirc_{n=0}^{\bar{N}-1} W^{[n]}\right)_{0,a-1} W^{[\bar{N}]}_{a-1,a} \nonumber\\
        &= \pihat{0}{a-1} \sum_{k=0}^{\bar{N}-1} \binom{\bar{N}-k-1}{a-1} \hat{Q}_k + \pihat{0}{a-2} \sum_{k=0}^{{\bar N}-1} \binom{\bar{N}-k-1}{a-2} \hat{Q}_k \sigma_{a-2}^{a-1} \nonumber \\
    &=  \pihat{0}{a-1} \sum_{k=0}^{\bar{N}-1}\left(  \binom{\bar{N}-k-1}{a-1} + \binom{\bar{N}-k-1}{a-2}\right) \hat{Q}_k \nonumber \\
    &=
    \pihat{0}{a-1} \sum_{k=0}^{\bar{N}} \binom{\bar{N}-k}{a-1} \hat{Q}_k,
\end{align}
as required, where in the final line, we use the fact that $\binom{\bar{N}-\bar{N}}{a-1} = 0 \; \forall a \in \mathbb{Z}^+$. Column $p+1$ of the top row equates to
\begin{align}
\left(\bigcirc_{n=0}^{{\bar N}} W^{[n]}\right)_{0,p+1}
&= W^{[{\bar N}]}_{0,p+1}
 + \sum_{k=1}^p \left( \left(\bigcirc_{n=0}^{{\bar N}-1} W^{[n]}\right)_{0,k} \, W^{[{\bar N}]}_{k,p+1} \right)
 + \left(\bigcirc_{n=0}^{{\bar N}-1} W^{[n]}\right)_{0,p+1} \nonumber\\
&= \hat{Q}_{\bar N}
 + \sum_{k=1}^p \left( \pihat{0}{k-1} \sum_{j=0}^{{\bar N}-1} \binom{{\bar N}-j-1}{k-1} \hat{Q}_j \, \sigma_{k-1}^p \right)
 + \sum_{k=0}^{{\bar N}-1} \hat{Q}_k \left( 1 + \sum_{m=1}^{p} \binom{{\bar N}-k-1}{m} \pihat{0}{m-1} \hat{\sigma}_{m-1}^p \right) \nonumber\\
&= \hat{Q}_{\bar N}
 + \sum_{j=0}^{{\bar N}-1} \sum_{k=1}^p \binom{{\bar N}-j-1}{k-1} \hat{Q}_j \, \pihat{0}{k-1} \sigma_{k-1}^p
 + \sum_{j=0}^{{\bar N}-1} \hat{Q}_j \left( 1 + \sum_{m=1}^{p} \binom{{\bar N}-j-1}{m} \pihat{0}{m-1} \hat{\sigma}_{m-1}^p \right) \nonumber\\
&= \hat{Q}_{\bar N}
 + \sum_{j=0}^{{\bar N}-1} \hat{Q}_j \left( \sum_{m=1}^{p} \binom{{\bar N}-j-1}{m-1} \pihat{0}{m-1} \sigma_{m-1}^p
 + 1 + \sum_{m=1}^{p} \binom{{\bar N}-j-1}{m} \pihat{0}{m-1} \hat{\sigma}_{m-1}^p \right) \nonumber\\
&= \hat{Q}_{\bar N}
 + \sum_{j=0}^{{\bar N}-1} \hat{Q}_j \left( 1 + \sum_{m=1}^{p} \bigl( \binom{{\bar N}-j-1}{m-1} + \binom{{\bar N}-j-1}{m} \bigr)
 \pihat{0}{m-1} \hat{\sigma}_{m-1}^p \right) \nonumber\\
&= \sum_{j=0}^{{\bar N}} \hat{Q}_j \left( 1 + \sum_{m=1}^{p} \binom{{\bar N}-j}{m} \pihat{0}{m-1} \hat{\sigma}_{m-1}^p \right).
\end{align}
as required. Finally, we reach the last column of row 0:
\begin{align}
\left(\bigcirc_{n=0}^{\bar N} W^{[n]}\right)_{0,p+2}
&= \left( \left(\bigcirc_{n=0}^{\bar N-1} W^{[n]}\right)_{0,1} W^{[\bar N]}_{1,p+2} \right)
 + \left( \left(\bigcirc_{n=0}^{\bar N-1} W^{[n]}\right)_{0,p+1} W^{[\bar N]}_{p+1,p+2} \right)
 + \left(\bigcirc_{n=0}^{\bar N-1} W^{[n]}\right)_{0,p+2} \nonumber\\
&= \sum_{k=0}^{\bar N-1} \hat{Q}_k \hat{Q}_{\bar N}
 + \sum_{k=0}^{\bar N-1} \hat{Q}_k \!\left( 1 + \sum_{m=1}^{p} \binom{\bar N-k-1}{m}\,\pihat{0}{m-1}\,\hat{\sigma}_{m-1}^{p} \right)\! \hat{Q}_{\bar N} \nonumber\\
&\quad + \sum_{k=1}^{\bar N-1} \sum_{l=0}^{k-1}
   \left( 2 \hat{Q}_{k} \hat{Q}_{l}
        + \sum_{s=1}^{p} \hat{Q}_{k}\hat{Q}_{l}\, \binom{k-l-1}{s}\,\pihat{0}{s-1}\,\hat{\sigma}_{s-1}^{p} \right) \nonumber\\
&= \sum_{l=0}^{\bar N-1} 2\,\hat{Q}_{\bar N}\hat{Q}_l
 + \sum_{l=0}^{\bar N-1}\sum_{s=1}^{p} \hat{Q}_{\bar N}\hat{Q}_l \,\binom{\bar N-l-1}{s}\,\pihat{0}{s-1}\,\hat{\sigma}_{s-1}^{p} \nonumber\\
&\quad + \sum_{k=1}^{\bar N-1} \sum_{l=0}^{k-1}
   \left( 2 \hat{Q}_{k} \hat{Q}_{l}
        + \sum_{s=1}^{p} \hat{Q}_{k}\hat{Q}_{l}\, \binom{k-l-1}{s}\,\pihat{0}{s-1}\,\hat{\sigma}_{s-1}^{p} \right) \nonumber\\
&= \sum_{k=1}^{\bar N} \sum_{l=0}^{k-1}
   \left( 2 \hat{Q}_{k} \hat{Q}_{l}
        + \sum_{s=1}^{p} \hat{Q}_{k}\hat{Q}_{l}\, \binom{k-l-1}{s}\,\pihat{0}{s-1}\,\hat{\sigma}_{s-1}^{p} \right).
\end{align}
as required.

For the first super-diagonal of our MPO, we exploit the upper-triangular form of our matrix to find
\begin{align}
    \left(\bigcirc_{n=0}^{\bar{N}} W^{[n]}\right)_{a,a+1} &= \left(\bigcirc_{n=0}^{\bar{N}-1} W^{[n]}\right)_{a,a+1} W^{[\bar{N}]}_{a+1,a+1} \nonumber\\ &+     \left(\bigcirc_{n=0}^{\bar{N}-1} W^{[n]}\right)_{a,a} W^{[\bar{N}]}_{a,a+1} \nonumber \\
    &= \left(\bigcirc_{n=0}^{\bar{N}-1} W^{[n]}\right)_{a,a+1} \mathbb{I} +      \mathbb{I} W^{[\bar{N}]}_{a,a+1}.
\end{align}
By definition, $W^{[\bar{N}]}_{a,a+1}$ is $\hat{Q}_{\bar{N}}$ for $a\in\{0,p+1\}$. Hence, for $a\in\{0,p+1\}$
\begin{align}
     \left(\bigcirc_{n=0}^{\bar{N}} W^{[n]}\right)_{a,a+1} &= \left(\bigcirc_{n=0}^{\bar{N}-1} W^{[n]}\right)_{a,a+1} + \hat{Q}_{\bar{N}} \nonumber \\
     &= \sum_{k=0}^{\bar{N}} \hat {Q}_k
\end{align}
as required. Similarly,  $W^{[\bar{N}]}_{a,a+1} = \sigma_{a-1}^{a}$ for $1 \leq a \leq p$, or $0$ otherwise. Therefore, for $1 \leq a \leq p$,
\begin{align}
     \left(\bigcirc_{n=0}^{\bar{N}} W^{[n]}\right)_{a,a+1} = \left(\bigcirc_{n=0}^{{\bar{N}}-1} W^{[n]}\right)_{a,a+1} + \hat\sigma_{a-1}^a \nonumber = \binom{{\bar{N}}}{1} \pihat{a-1}{a} + \hat{\sigma}_{a-1}^{a} = \binom{N}{1} \pihat{a-1}{a}.
\end{align}
More generally, the $k$th upper superdiagonal is given for $\leq a + k \leq p$ as
\begin{align}
    \left(\bigcirc_{n=0}^{\bar{N}} W^{[n]}\right)_{a,a+k} &=      \sum_{j=a}^{a+k} \left(\bigcirc_{n=0}^{\bar{N}-1} W^{[n]}\right)_{a,j} W^{[\bar{N}]}_{j,a+k} \nonumber \\
    &= \left(\bigcirc_{n=0}^{\bar{N}-1} W^{[n]}\right)_{a,a+k}  + \left(\bigcirc_{n=0}^{\bar{N}-1} W^{[n]}\right)_{a,a+k-1}
    W^{[\bar{N}]}_{a+k-1,a+k} \nonumber \\
    &= \binom{\bar{N}}{k} \pihat{a-1}{a+k-1}  + \binom{\bar{N}}{k-1} \pihat{a-1}{a+k-2}
    \sigma_{a+k-2}^{a+k-1}\nonumber \\
    &= \left(\bigcirc_{n=0}^{\bar{N}-1} W^{[n]}\right)_{a,a+k}  + \left(\bigcirc_{n=0}^{\bar{N}-1} W^{[n]}\right)_{a,a+k-1}
    W^{[\bar{N}]}_{a+k-1,a+k} \nonumber \\
    &= \binom{\bar{N}}{k} \pihat{a-1}{a+k-1}  + \binom{\bar{N}}{k-1} \pihat{a-1}{a+k-1}
   \nonumber \\
   &= \binom{N}{k} \pihat{a-1}{a+k-1},
\end{align}
which aligns with Equation \eqref{polynomialMPO_Nn}.
Having verified the top row and all columns up to and including $p$, we now turn to rows $1$ to $p$ of column $p+1$:
\begin{align}
    \left(\bigcirc_{n=0}^{{\bar{N}}} W^{[n]}\right)_{a,p+1} &=
    \sum_{j=a}^{p+1} \left(\bigcirc_{n=0}^{{\bar{N}}-1} W^{[n]}\right)_{a,j} W^{[{\bar{N}}]}_{j,p+1} \nonumber \\
    &=\sum_{j=a}^p \left(\bigcirc_{n=0}^{{\bar{N}}-1} W^{[n]}\right)_{a,j} W^{[\bar{N}]}_{j,p+1} + \left(\bigcirc_{n=0}^{{\bar{N}}-1} W^{[n]}\right)_{a,p+1} \nonumber \\
    &= \sum_{j=a}^p
    \binom{{\bar{N}}}{j-a} \pihat{a-1}{j-1}
     \hat\sigma_{j-1}^{p}  + \sum_{k=0}^{p-a} \binom{{\bar{N}}}{k+1}\pihat{a-1}{k+a-1} \hat{\sigma}_{k+a-1}^p
     \nonumber \\
    &= \sum_{k=0}^{p-a}
    \binom{{\bar{N}}}{k} \pihat{a-1}{k+a-1}
     \hat\sigma_{k+a-1}^{p}  + \sum_{k=0}^{p-a} \binom{{\bar{N}}}{k+1}\pihat{a-1}{k+a-1} \hat{\sigma}_{k+a-1}^p
     \nonumber \\
    &= \sum_{k=0}^{p-a}
\left(    \binom{{\bar{N}}}{k}
     +\binom{{\bar{N}}}{k+1}\right) \pihat{a-1}{k+a-1} \hat{\sigma}_{k+a-1}^p \nonumber \\
    &= \sum_{k=0}^{p-a}
\binom{N}{k+1}
\pihat{a-1}{k+a-1} \hat{\sigma}_{k+a-1}^p,
\end{align}
as required.

Finally, we complete the inductive proof by verifying the final column, $p+2$. We have already verified the top row of column $p+2$. For row $1$, we get:
\begin{align}
    \left(\bigcirc_{n=0}^{{\bar{N}}} W^{[n]}\right)_{1,p+2} &=
    \sum_{j=1}^{p+2} \left(\bigcirc_{n=0}^{{\bar{N}}-1} W^{[n]}\right)_{1,j} W^{[{\bar{N}}]}_{j,p+2} \nonumber \\
    &= W^{[{\bar{N}}]}_{1,p+2} + \left(\bigcirc_{n=0}^{{\bar{N}}-1} W^{[n]}\right)_{1,p+1} W^{[{\bar{N}}]}_{p+1,p+2} + \left(\bigcirc_{n=0}^{{\bar{N}}-1} W^{[n]}\right)_{1,p+2} \nonumber \\
    &= \hat{Q}_{\bar{N}} + \sum_{k=0}^{p-1} \binom{{\bar{N}}}{k+1} \pihat{0}{k} \hat \sigma_{k}^{p} \hat{Q}_{\bar{N}} +
    \sum_{k=0}^{{\bar{N}}-1} \hat{Q}_k \left(1+ \sum_{m=1}^p \binom{k}{m} \pihat{0}{m-1} \hat\sigma_{m-1}^{p}\right)
     \nonumber \\
     &= \hat{Q}_{\bar{N}} \left(1 + \sum_{k=0}^{p-1} \binom{{\bar{N}}}{k+1} \pihat{0}{k} \hat \sigma_{k}^{p} \right) +
    \sum_{k=0}^{{\bar{N}}-1} \hat{Q}_k \left(1+ \sum_{m=1}^{p} \binom{k}{m} \pihat{0}{m-1} \hat\sigma_{m-1}^{p}\right)
     \nonumber \\
    &= \hat{Q}_{\bar{N}} \left(1 + \sum_{m=1}^{p} \binom{{\bar{N}}}{m} \pihat{0}{m-1} \hat \sigma_{m-1}^{p} \right) +
    \sum_{k=0}^{{\bar{N}}-1} \hat{Q}_k \left(1+ \sum_{m=1}^p \binom{k}{m} \pihat{0}{m-1} \hat\sigma_{m-1}^{p}\right)
     \nonumber \\
     &=\sum_{k=0}^{{\bar{N}}} \hat{Q}_k \left(1+ \sum_{m=1}^p \binom{k}{m} \pihat{0}{m-1} \hat\sigma_{m-1}^{p}\right),
\end{align}
as required. For rows $2$ to $p+1$, we have
\begin{align}
    \left(\bigcirc_{n=0}^{{\bar{N}}} W^{[n]}\right)_{a,p+2} &=
    \sum_{j=a}^{p+2} \left(\bigcirc_{n=0}^{{\bar{N}}-1} W^{[n]}\right)_{a,j} W^{[{\bar{N}}]}_{j,p+2} \nonumber \\
    &= \left(\bigcirc_{n=0}^{{\bar{N}}-1} W^{[n]}\right)_{a,p+1} W^{[{\bar{N}}]}_{p+1,p+2} + \left(\bigcirc_{n=0}^{{\bar{N}}-1} W^{[n]}\right)_{a,p+2} \nonumber \\
&= \sum_{k=0}^{p-a}\binom{{\bar{N}}}{k+1} \pihat{a-1}{k+a-1} \hat\sigma_{k+a-1}^p \hat{Q}_{\bar{N}} +
\sum_{k=0}^{{\bar{N}}-1} \hat{Q}_k \sum_{m=1}^{p-a+1} \binom{k}{m} \pihat{a-1}{m+a-2} \hat \sigma_{m+a-2}^{p} \nonumber \\
&= \hat{Q}_{\bar{N}} \sum_{m=1}^{p-a+1}\binom{{\bar{N}}}{m} \pihat{a-1}{m+a-2} \hat\sigma_{m+a-2}^p  +
\sum_{k=0}^{{\bar{N}}-1} \hat{Q}_k \sum_{m=1}^{p-a+1} \binom{k}{m} \pihat{a-1}{m+a-2} \hat \sigma_{m+a-2}^{p} \nonumber \\
&= \sum_{k=0}^{{\bar{N}}} \hat{Q}_k \sum_{m=1}^{p-a+1} \binom{k}{m} \pihat{a-1}{m+a-2} \hat \sigma_{m+a-2}^{p},
\end{align}
as required. Finally, for row $p+1$ of column $p+2$, we see
\begin{align}
    \left(\bigcirc_{n=0}^{{\bar{N}}} W^{[n]}\right)_{p+1,p+2} &= \left(\bigcirc_{n=0}^{{\bar{N}}-1} W^{[n]}\right)_{p+1,p+1} W^{[{\bar{N}}]}_{p+1,p+2} + \left(\bigcirc_{n=0}^{{\bar{N}}-1} W^{[n]}\right)_{p+1,p+2} \nonumber \\
&= \hat{Q}_{\bar{N}} + \sum_{k}^{{\bar{N}}-1} \hat{Q}_k \nonumber \\
&= \sum_{k}^{{\bar{N}}} \hat{Q}_k,
\end{align}
completing the proof by induction. \qed
\end{widetext}

\end{document}